\documentclass[aps,prb,reprint,amsfonts,amsmath,amssymb]{revtex4-1}
\usepackage[colorlinks=true,citecolor=blue,linkcolor=blue,filecolor=blue,urlcolor=blue]{hyperref}
\usepackage{bm,braket}
\usepackage[pdftex]{graphicx}
\begin{document}

%%%%%%%%%%%%%%%%%%%%%%%%%%%%%%%%%%%%%%%%%%%%%%%%%%%%%%%%%%%%
%%%%%%%%%%%%%%% title and author information %%%%%%%%%%%%%%%
%%%%%%%%%%%%%%%%%%%%%%%%%%%%%%%%%%%%%%%%%%%%%%%%%%%%%%%%%%%%
\title{Phase diagram of the square lattice Hubbard model with Rashba-type antisymmetric spin-orbit coupling}
\author{Masataka Kawano$^{1,2}$}
\email{masataka.kawano@tum.de}
\author{Chisa Hotta$^{2}$}
\affiliation{$^{1}$Department of Physics, Technical University of Munich, 85748 Garching, Germany\\
$^{2}$Department of Basic Science, University of Tokyo, Meguro-ku, Tokyo 153-8902, Japan}
\date{\today}

%%%%%%%%%%%%%%%%%%%%%%%%%%%%%%%%%%%%%%%%
%%%%%%%%%%%%%%% abstract %%%%%%%%%%%%%%%
%%%%%%%%%%%%%%%%%%%%%%%%%%%%%%%%%%%%%%%%
\begin{abstract}
We clarify the ground state phase diagram of the half-filled 
square-lattice Hubbard model with Rashba spin-orbit coupling (SOC) 
characterized by the spin-split energy bands due to broken inversion symmetry. 
Although the  Rashba metals and insulating magnets have been studied well, 
%%the metallic and strong coupling limits of this model have been studied well, 
the intermediate interaction strength of the system remained elusive due to the lack of appropriate theoretical tools to 
unbiasedly describe the large-scale magnetic structures. 
We complementarily apply four different methods; sine-square deformed mean-field theory,
random phase approximation(RPA), Luttinger-Tisza method, and density matrix embedding theory,
and succeed in capturing the incommensurate spin-density-wave (SDW) phases with very long spatial periods 
which were previously overlooked. 
The transition to the SDW phases from the metallic phase is driven by an unprecedented instability 
that nests the two parts of the Fermi surface carrying opposite spins. 
For large SOC, the spiral, stipe, and vortex phases are obtained, 
when the four Dirac points exist near the Fermi level and their whole linear dispersions nest 
by a wavelength $\pi$, opening a band gap. 
These two types of transition provide Fermiology that distinguishes the antisymmetric SOC systems, 
generating a variety of magnetic phases that start from the relatively weak correlation regime 
and continue to the strongly interacting limit. 
\end{abstract}
\maketitle

%%%%%%%%%%%%%%%%%%%%%%%%%%%%%%%%%%%%%%%
%%%%%%%%%%%%%%% section %%%%%%%%%%%%%%%
%%%%%%%%%%%%%%%%%%%%%%%%%%%%%%%%%%%%%%%
\section{Introduction}
Spin-orbit coupling (SOC) plays a crucial role in controlling spin-dependent transport phenomena~\cite{nagaosa2010rmp,sinova2015rmp},
in the emergence of topological phases of matter~\cite{qi2011rmp}, 
and in stabilizing a variety of magnetic structures of insulators~\cite{togawa2016jpsj,nagaosa2013nnano}.
Intensive studies over the past decades have clarified many of these key roles, 
and it turned out that there is an inherent difference in how the SOC works for crystals \textit{with and without inversion symmetry} as summarized in Table~\ref{fig:chart}. 
\par
The so-called non-centrosymmetric crystals do not have inversion symmetry, 
and its SOC becomes antisymmetric about left-moving and right-moving electrons on each bond. 
Accordingly, the electron spin momentum couples with the kinetic momentum $\bm{k}$, and the energy bands split, 
carrying spins pointing in the directions that vary with $\bm{k}$. 
In particular, the Rashba or Dresselhaus types of spin-split bands in two dimensions have given rise to
a variety of spin-dependent transport phenomena
including spin Hall effect~\cite{dyakonov1971,hirsch1999,murakami2003,sinova2004,inoue2004,kato2004,wunderich2005}
and spin galvanic effect~\cite{edelstein1990ssc,kato2004prl,silov2004apl}. 
When the system becomes insulating, the antisymmetric SOC is converted to 
the Dzyaloshinskii-Morita (DM) interaction~\cite{dzyaloshinskii1958jpcs,moriya1960pr}. 
The DM interaction typically competes with the Heisenberg exchange interaction, 
resulting in magnetic orders with spatially extended periods 
such as chiral magnets in one dimension (1D)~\cite{dzyaloshinskii1964spj,dzyaloshinskii1965spj,moriya1982ssc,miyadai1983jpsj,togawa2012prl} and magnetic skyrmions in two dimensions (2D)~\cite{robler2006nature,muhlbauer2009science,yu2010nature,heinze2011natphys}. 
Even when the DM interactions appear to be irrelevant, low-energy magnon excitations 
are well-influenced and reveals the nonreciprocal magnon propagation~\cite{melcher1973prl,kataoka1987jpsj,zakeri2010prl,iguchi2015prb,gitgeatpong2017prl,iguchi2018prb,cheon2018prb},
and spin-dependent magnon-band splitting \cite{okuma2017prl,kawano2019cp,kawano2019prb2}.
\par
When the inversion symmetry in the centrosymmetric crystals coexists with the time-reversal symmetry, 
the entire energy band retains its spin-degeneracy.
In $4d$ and $5d$ materials, the strong SOC overwhelms the energy scale of the crystal field, 
and combines electron spin and orbital angular momentum and form a Kramers doublet, 
which is the origin of the band-degeneracy. 
When these materials have a valence that fills these doublets by half, 
the strong Coulomb interaction often drives the system to the so-called 
``spin-orbit coupled Mott insulators''. 
There, the SOC produces an unusual distribution of the spin density in real space~\cite{kim2008prl,kim2009science,jackeli2009} and serves as a source of various intriguing quantum phases including 
quantum spin liquid phases~\cite{kitaev2006,jackeli2009,chaloupka2010,singh2010,singh2012,plumb2014,kubota2015},
topological Mott insulator~\cite{pesin2010np},
Weyl semimetals~\cite{murakami2007njp,wan2011prb},
multipolar phases~\cite{chen2010prb,chen2011prb},
and the perfect flat bands and trimerized charge orderings driven by SOC~\cite{nakai2021ncom}.
%%%%%%%%%%%%%%%%%%%%%%%%%%%%%%%%%%%%%%%%%%%%%%%%%%%%%%%%%%%%%%%%%%%%%%%%%%%%%%%%%%%%%%%%%%%%%%%%%%%%%%%%%%%%%%%%%%%%%%%%
\begin{table}
\caption{Classification of SOC systems. The region marked with a broken line is the target of the present paper.}
\includegraphics[width=85mm]{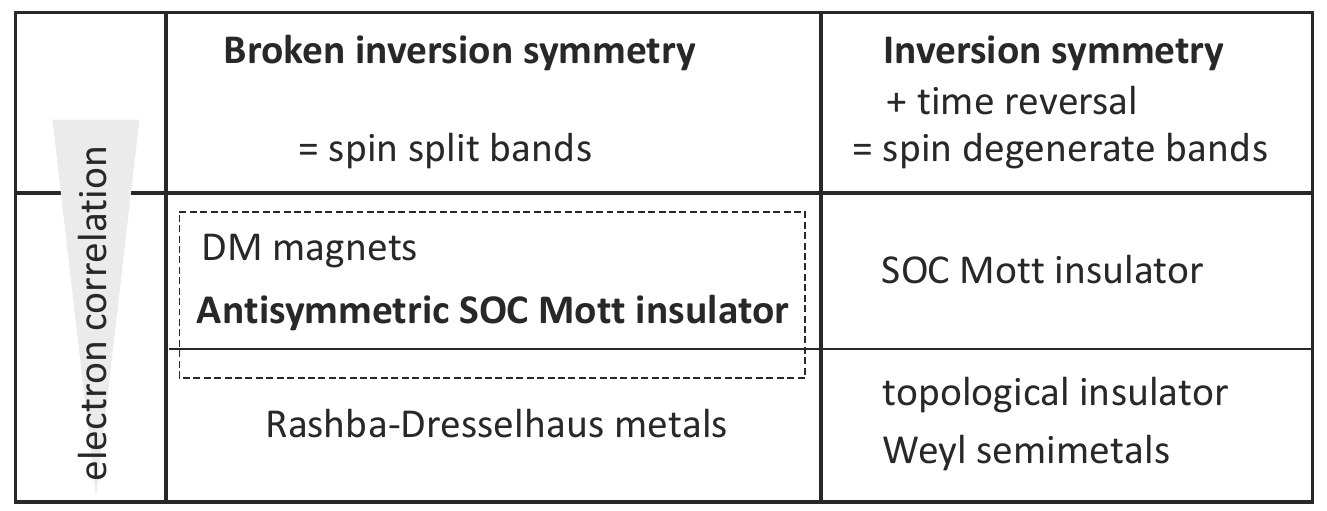}
\label{fig:chart}
\end{table}
%%%%%%%%%%%%%%%%%%%%%%%%%%%%%%%%%%%%%%%%%%%%%%%%%%%%%%%%%%%%%%%%%%%%%%%%%%%%%%%%%%%%%%%%%%%%%%%%%%%%%%%%%%%%%%%%%%%%%%%%
\par
These studies show that the role of antisymmetric SOC and symmetric SOC are entirely different.
The former explicitly makes the energy bands spin-momentum-dependent, 
while in the latter, the spin degrees of freedom are masked in the bulk by the degeneracy 
and the SOC rather works to mix different parity of the localized basis, 
adding topological nature to the energy bands which manifests in the spin-oriented edge states. 
\par
Here, we notice that the antisymmetric SOC for intermediate correlation strength 
in the inversion-symmetry-broken cases is essentially unexplored, 
in comparison to the established and still increasing fields of 
the SOC Mott insulator with inversion symmetry. 
We call them {\it ``antisymmetric SOC Mott insulators''}. 
The reason, from a theoretical standpoint, can be explained as a lack of appropriate tools. 
The antisymmetric SOC is expected to yield quantum phases with a large-scale structure. 
However, dealing with such phases in the presence of strong electronic interactions is extremely difficult
due to their long-period structure;
all methods known so far require prior knowledge of the structure or periods of ordered phases,
and the results are too often dependent on the numerical conditions~\cite{shibata2011prb,hotta2012prb,nishimoto2013ncom,kawano2022prr}. 
Previous theories have thus focused on the non-interacting Rashba-Dresselhaus types of metallic phases,
superconducting phases mostly at the mean-field level with perturbations~\cite{yanase2008jpsj,yanase2013jpsj,Greco2018,Greco2020,wolf2020prb,nogaki2020prb,nogaki2021prr,nogaki2022prb,beyer2022arxiv},
and on the insulating state in the strong coupling limit by further simplifying them as classical magnets
to avoid the quantum many-body effect.
%Previous theories have thus focused either on the non-interacting Rashba-Dresselhaus types of metallic phases 
%or on the insulating state in the strong coupling limit by further simplifying them as classical magnets 
%to avoid the quantum many-body effect. 
\par
On the experimental side, 
until recently, only a few materials with substantial antisymmetric SOC and strong electron correlation 
had been identified experimentally. 
Meanwhile, a few $3d$ electron systems started to be highlighted~\cite{sunko2017nature}, 
and the antisymmetric SOC is artificially introduced in cold atomic system~\cite{lin2011nature,zhang2012prl,wang2012prl,cheuk2012prl},
and the theory dealing with the antisymmetric Mott insulator is demanding. 
\par
In this paper, we address this issue using the simplest prototype platform, 
the square-lattice Hubbard model with Rashba SOC. 
Since the Hamiltonian is expected to host a series of intriguing magnetic structures of a large spatial scale, 
we apply a numerical method we developed very recently, the sine-square deformed mean-field theory (SSDMF)~\cite{kawano2022prr}, 
and combine them with the analysis from the weak coupling random-phase approximation (RPA) 
strong coupling Luttinger-Tisza method, 
and the density matrix embedding theory (DMET)~\cite{knizia2012prl,knizia2013jctc,wouters2016jctc}. 
\par
The ground state phase diagram is obtained from weak to strong SOC and Coulomb interactions using the SSDMF, 
which shows three spin-density-wave (SDW) phases with modulated spin amplitude
as well as spiral, stripe, and vortex phases with spatially rotating spins of equal amplitudes. 
The RPA analysis reveals several insights into the mechanism of the metal-to-magnetic phase transitions; 
the spin-split bands show the spin-selective Fermi-surface nesting instability, 
which is responsible for the formation of incommensurate SDWs. 
Whereas, we find that there are four Dirac cones and 
the entire linear dispersions of two of them ``nests'' to the other two via a commensurate wave-vector 
and open a bandgap, which yields the stripe and vortex phases. 
\par
From the methodological point of view, the obtained phase diagram serves as a benchmark of the SSDMF, 
demonstrating that it can accurately determine all possible types of large-scale spatial periods of magnetic orderings. 
The SSDMF was previously tested and confirmed in the 1D and 2D models having reference solutions, 
and this work is the first application to the unexplored problem. 
To verify the results of the SSDMF, we apply the DMET, which takes account of the full electronic correlation effects, 
and at the same time, properly expresses the incommensurate phases of large-scale periods. 
We finally point out that a gauge invariant Wilson loop can explain parts of the phase boundaries, 
and can be used to prove that the quantum spin liquid (QSL) phase previously reported does not exist in the phase diagram. 
\par
This paper is organized as follows.
In Sec.~\ref{sec:model&phase}, 
we introduce the model and give the overview of the results and the brief explantions on their physical implications. 
There, one finds the information on among which part of the following subsections 
in Sec.\ref{sec:weak-coupling}-\ref{sec:dmet}, the detail will be found; 
these subsections are made independent and can be chosen for one's purposes. 
In Sec.~\ref{sec:weak-coupling} and Sec.~\ref{sec:strong-coupling},
we clarify the magnetic phases in weak- and strong-coupling limits. 
In Sec.~\ref{sec:ssdmf} and Sec.~\ref{sec:dmet}, 
we apply the SSDMF and DMET to our model. 
In these sections, we provide minimal explanations of the methods 
to understand the physical implication of our results, 
where we refer to the points that are specific to our model. 
Those who want to learn more about the latter two methods consistently
shall visit e.g. Refs. [\onlinecite{kawano2022prr}] and 
[\onlinecite{kawano2020prb}] by the authors. 
In Sec.~\ref{sec:wilson}, 
we discuss the relationship between gauge-invariant Wilson loops and magnetic phase boundary.
We finally give a brief discussion and summary in Secs.~\ref{sec:discussion} and \ref{sec:summary}.

%%%%%%%%%%%%%%%%%%%%%%%%%%%%%%%%%%%%%%%
%%%%%%%%%%%%%%% section %%%%%%%%%%%%%%%
%%%%%%%%%%%%%%%%%%%%%%%%%%%%%%%%%%%%%%%
\section{Model and magnetic phase diagram}
\label{sec:model&phase}
%%%%%%%%%% Subsection %%%%%%%%%%
\subsection{Model}

\begin{figure}[t]
\includegraphics[width=85mm]{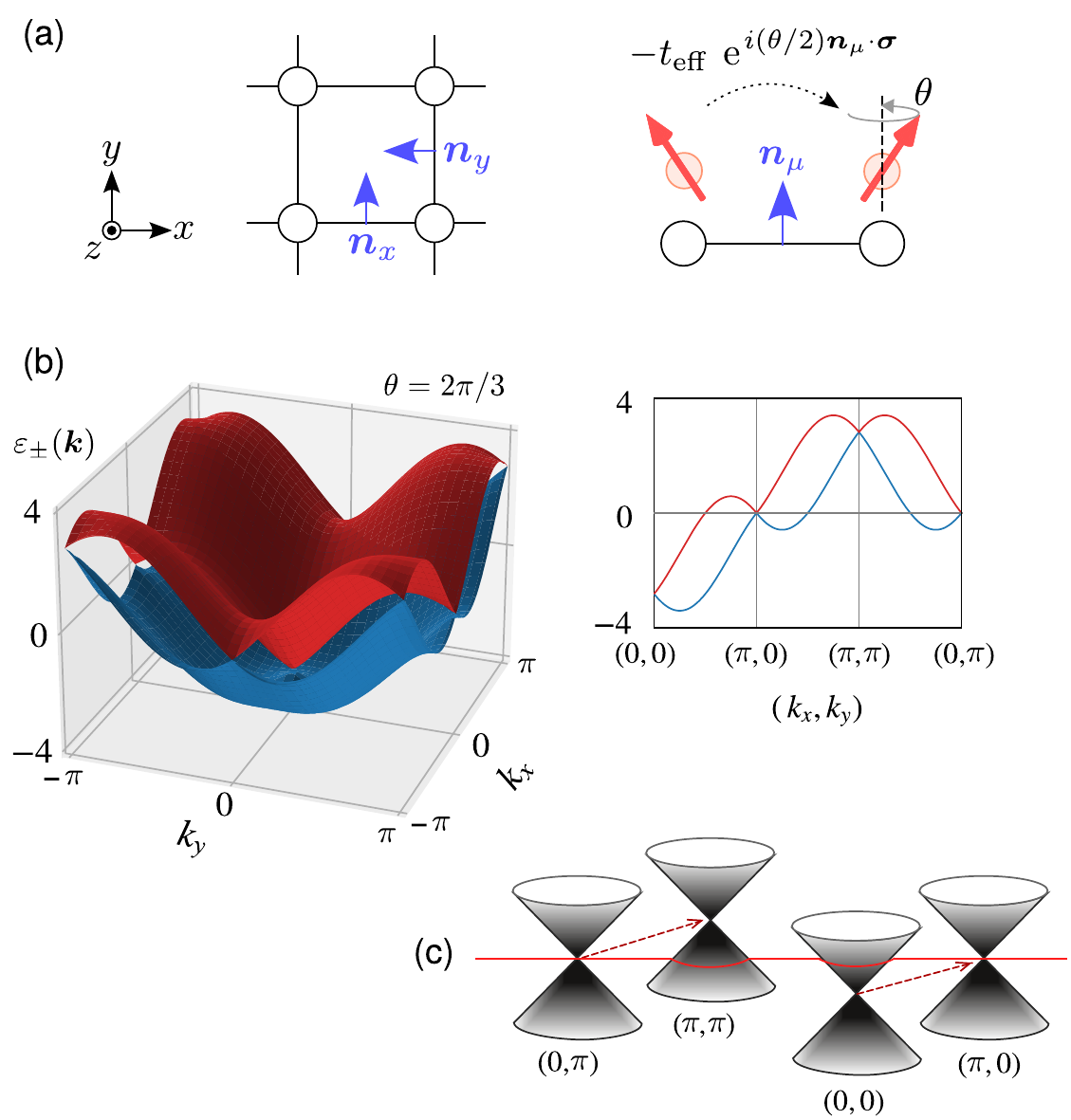}
\caption{(a) 2D square lattice with Rashba SOC in Eq.~(\ref{eq:model}), where $\bm{n}_\mu$ ($\mu=x,y$) denotes the 
unit vector that rotates the spins in hopping to its neighbors by the SU(2) gauge field shown 
schematically in the right panel. 
(b) Representative non-interacting energy band structure $\varepsilon_\pm(\bm k)$ of $\hat{\mathcal{H}}_{0}$ 
for $\theta=2\pi/3$. The left panel is the one along the symmetric lines. 
(c) Schematic illustration of the four Dirac cones with the same velocities. 
The broken arrows represent the $\bm q=(0,\pi), (\pi,0)$ vector that ``nests'' the whole Dirac cones 
in forming the stripe and vortex phases. 
}
\label{fig:model}
\end{figure}

We consider the half-filled Hubbard model with Rashba SOC defined on the square lattice of system size $N=L\times L$, 
given as $\hat{\mathcal{H}}=\hat{\mathcal{H}}_{0}+\hat{\mathcal{H}}_{U}$, with 
%%%%%%%%%%%%%%%
\begin{align}
\hat{\mathcal{H}}_{0}
&=
-
\sum_{\bm{r}}
\sum_{\mu=x,y}
(
t
\hat{\bm{c}}_{\bm{r}+\bm{e}_{\mu}}^{\dagger}
\hat{\bm{c}}_{\bm{r}}
+i\lambda
\hat{\bm{c}}_{\bm{r}+\bm{e}_{\mu}}^{\dagger} 
(\bm{n}_{\mu}\cdot\bm{\sigma})
\hat{\bm{c}}_{\bm{r}}
+
\mathrm{h.c.}
)
,
\\
\hat{\mathcal{H}}_{U}
&=
U
\sum_{\bm{r}}
\left(
\hat{n}_{\bm{r},\uparrow}
-
\frac{1}{2}
\right)
\left(
\hat{n}_{\bm{r},\downarrow}
-
\frac{1}{2}
\right)
,
\end{align}
%%%%%%%%%%%%%%%
where $\bm{r}$ denotes the position of sites on the square lattice,
$\bm{e}_{x}=(1,0)$ and $\bm{e}_{y}=(0,1)$ are the unit vectors connecting nearest-neighboring sites
\footnote{The susbscipt (superscript) like $\bm{e}_{\mu}$ ($\bm{e}^{\mu}$) denote the direction in real (spin) space.},
$\hat{c}_{\bm{r},\sigma}$ ($\hat{c}_{\bm{r},\sigma}^{\dagger}$)
in $\hat{\bm{c}}_{\bm{r}}=(\hat{c}_{\bm{r},\uparrow},\hat{c}_{\bm{r},\downarrow})^{T}$
denotes the annihilation (creation) operator of the electron at site $\bm{r}$ with spin $\sigma=\uparrow,\downarrow$,
and $\hat{n}_{\bm{r},\sigma}=\hat{c}_{\bm{r},\sigma}^{\dagger}\hat{c}_{\bm{r},\sigma}$ denotes the particle density with spin $\sigma$. 
The electrons interact via the on-site repulsive interaction, $U\geq0$, 
where we introduce the factor $1/2$ in $\hat{\mathcal{H}}_{U}$ to recover the particle-hole symmetry.
The schematic illustration of the model is shown in Fig.~\ref{fig:model}(a).
%%%%%%%%%%%%%%%%%%%%%%%%%%%%%%%%%%%%%%%%%%%%%%%%%%%%%%%%%%%%%%%%%%%%%%%%%%%%%%%%%%%%%%%%%%%%%%%%%%%%%%%%%%%%%%%%%%%%%%%%
%%%%%%%%%%%%%%%%%%%%%%%%%%%%%%%%%%%%%%%%%%%%%%%%%%%%%%%%%%%%%%%%%%%%%%%%%%%%%%%%%%%%%%%%%%%%%%%%%%%%%%%%%%%%%%%%%%%%%%%%
\par
The non-interacting Hamiltonian $\hat{\mathcal{H}}_{0}$ includes the standard hopping integral $t$
and spin-dependent hopping integral $\lambda$ that originates from the Rashba SOC.
These two hopping terms are combined and are rewritten as,
%%%%%%%%%%%%%%%
\begin{align}
\hat{\mathcal{H}}_{0}
&=
-t_{\mathrm{eff}}
\sum_{\bm{r}}
\sum_{\mu}
\left(
\hat{\bm{c}}_{\bm{r}+\bm{e}_{\mu}}^{\dagger}
\mathrm{e}^{i(\theta/2)\bm{n}_{\mu}\cdot\bm{\sigma}}
\hat{\bm{c}}_{\bm{r}}
+
\mathrm{h.c.}
\right)
,
\label{eq:model}
\end{align}
%%%%%%%%%%%%%%%
where $t_{\mathrm{eff}}=\sqrt{t^{2}+\lambda^{2}}$,
$\theta=2\mathrm{arctan}(\lambda/t)$,
$\bm{n}_{x}=(0,1,0)$, $\bm{n}_{y}=(-1,0,0)$,
$\bm{\sigma}=(\sigma^{x},\sigma^{y},\sigma^{z})$ is the Pauli matrix,
and we take $t_{\mathrm{eff}}=1$ in the following.
The $2\times2$ matrix $\mathrm{e}^{i(\theta/2)\bm{n}_{\mu}\cdot\bm{\sigma}}$
can be regarded as an SU(2) gauge field~\cite{frohlich1993rmp},
which rotates the electron spin by $\theta$ about the unit vector $\bm{n}_{\mu}$
when the electron hops between the two sites as shown in Fig.~\ref{fig:model}(a).
\par
We confine ourselves to the region $0\leq\theta\leq\pi$
since the results for the other parameters can be obtained by applying a unitary transformation
$\hat{\mathcal{U}}\hat{\mathcal{H}}\hat{\mathcal{U}}^{\dagger}$, where
%%%%%%%%%%%%%%%
\begin{align}
\hat{\mathcal{U}}
=
\bigotimes_{\bm{r}}
\left(
\bigotimes_{\sigma}
(
\hat{c}_{\bm{r},\sigma}
-
\hat{c}_{\bm{r},\sigma}^{\dagger}
)
\right)
\exp
\left(
-i\pi
\hat{S}_{\bm{r}}^{x}
\right)
,
\end{align}
%%%%%%%%%%%%%%%
and $\hat{S}_{\bm{r}}^{\mu}=(1/2)\hat{\bm{c}}_{\bm{r}}^{\dagger}\sigma^{\mu}\hat{\bm{c}}_{\bm{r}}$ ($\mu=x,y,z)$
is the spin operator.
This unitary operator transforms $\theta$ into $2\pi-\theta$ in the Hamiltonian.
Therefore,
once one obtains the results in the region $0\leq\theta\leq\pi$,
those of $\pi\leq\theta\leq2\pi$ are automatically derived by applying $\hat{\mathcal{U}}$,
and repeating this operation gives results for all parameter regions $0\leq\theta\leq4\pi$.
We also remark that Rashba SOC in this model can be transformed to Dresselhaus SOC by a unitary operator
%%%%%%%%%%%%%%%
\begin{align}
\hat{\mathcal{U}}_{\mathrm{RD}}
=
\bigotimes_{\bm{r}}
\exp
\left(
i\pi
\frac{-\hat{S}_{\bm{r}}^{x}+\hat{S}_{\bm{r}}^{y}}{\sqrt{2}}
\right)
,
\end{align}
%%%%%%%%%%%%%%%
which represents the global $\pi$ rotation of the spin about the unit vector $(-\bm{e}^{x}+\bm{e}^{y})/\sqrt{2}$.
Then all the results obtained in this paper can also be applied to the square-lattice Hubbard model with Dresselhaus SOC;
the only difference between the two models is the direction of spins.
%%%%%%%%%%%%%%%%%%%%%%%%%%%%%%%%%%%%%%%%%%%%%%%%%%%%%%%%%%%%%%%%%%%%%%%%%%%%%%%%%%%%%%%%%%%%%%%%%%%%%%%%%%%%%%%%%%%%%%%%
\begin{figure*}[t]
\includegraphics[width=175mm]{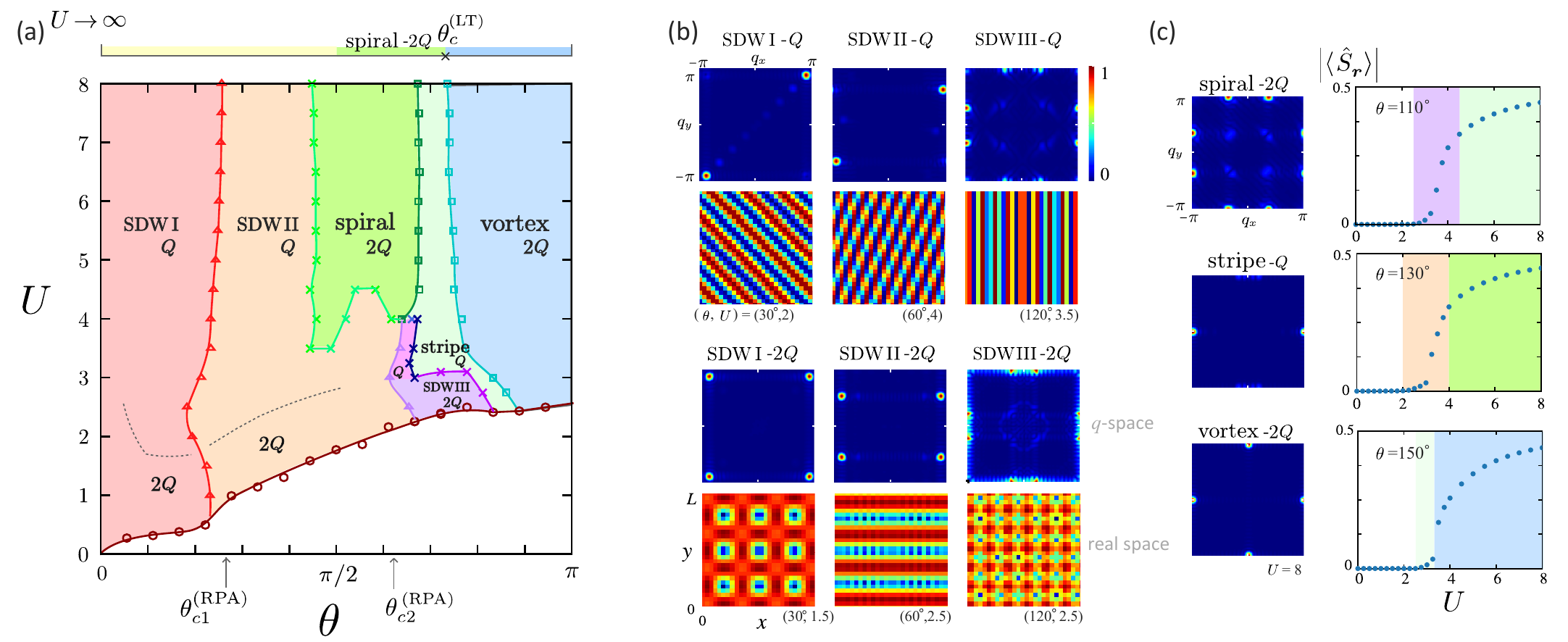}
\caption{(a) Ground state magnetic phase diagram obtained by the SSDMF with $L=28$.
There are three incommensurate spin-density-wave (SDW) orders labeled by SDWI, SDWII, and SDWIII.
For large $\theta$, the system hosts the spiral, stripe, and vortex order. 
(b) Spin structure factor (upper panels) and the spin density distribution in real space (lower panels) 
given as density plots, normalizing the minimum to maximum values as $[0:1]$ 
for the SDW phases with single- and double-$\bm Q$ (denoted as $Q$ and $2Q$). 
(c) Spin structure factor and the spin density as a function of $U$ for spiral, stripe, and vortex phases. 
}
\label{fig:phase}
\end{figure*}
%%%%%%%%%%%%%%%%%%%%%%%%%%%%%%%%%%%%%%%%%%%%%%%%%%%%%%%%%%%%%%%%%%%%%%%%%%%%%%%%%%%%%%%%%%%%%%%%%%%%%%%%%%%%%%%%%%%%%%%%

%%%%%%%%%% Subsection %%%%%%%%%%
\subsection{Magnetic phase diagram}

Figure~\ref{fig:phase}(a) shows the ground-state magnetic phase diagram of our model on the plane of $\theta$ and $U$. 
It is obtained by the SSDMF method which can unbiasedly capture the existing long-wavelength incommensurate orderings 
and properly judge their stability, as explained in Sec.~\ref{sec:ssdmf}. 
The SSDMF does not formally/basically include correlation effect beyond mean-field approximation,
however, possibly because of the particular nature of their wave function,
it is previously shown to describe the correlated model beyond the mean field~\cite{kawano2022prr}.
We find five distinct magnetically-ordered phases labeled as SDWI, SDWII, SDWIII, spiral, stripe, and vortex phases.
At $\theta=0$ (without SOC), an infinitesimal $U$ induces an antiferromagnetic (AFM) order. 
When $\theta$ takes a small nonzero value, 
the SDWI immediately appears. 
At $\theta \sim \pi/5$, it transforms to the SDWII order and to the SDWIII order at $\theta\sim 2\pi/3$, 
and at larger $\theta$ and $U$, we find spiral, stripe, and vortex phases. 
\par
The structure factors of all the phases that appear in the phase diagram are shown in 
Figs.~\ref{fig:phase}(b) and \ref{fig:phase}(c). 
The peak positions in the SDWI phase are at the incommensurate $q_x=q_y$ point, 
and they shift to $q_x\ne q_y$ for SDWII and in SDWIII to the boundary (off the symmetric points) 
of the Brillouin zone. The $\bm q$-profile of SDWIII is similar to the stripe phase, 
and in the spiral and vortex phases, $q_x,q_y$ take $0$ or $\pi$.
These results indicate that the ordering patterns gradually develop
from the incommensurate to the commensurate ones as $\theta$ increases.
In the weaker-$U$ region, we find the double-$\bm Q$ structure in the SDW phases, 
which is detected as the doubling of the peaks. 
This should be because the magnetic structures with double peaks will increase the splitting of the energy bands 
at the phase transition. 
However,
the amplitudes of spin-moments of the SDW in real space
have larger variations for the double-$\bm Q$ structures, 
and the nonmagnetic sites or sites with small spin moments 
increase compared to the single-$\bm Q$ profile. 
We have checked that the degree of double occupancy is larger for the double-$\bm{Q}$ structure, which is 
unfavorable for $U$, which is the reason why the double-$\bm Q$ is replaced 
to the single-$\bm Q$ SDW phases as $U$ increases.
The single-$\bm{Q}$ SDW phases have a coplanar spin structure by definition, 
and the double-$\bm{Q}$ SDW phases have non-coplanar one, 
reflecting the large spatial variation of the spin moments.
\par
In the spiral, stripe, and vortex phases, the spin density no longer varies in space, 
and the spin orientation rotates with a commensurate periodicity. 
When compared to the average spin densities of the SDW phases, 
they take $|\langle S_{\bm{r}}\rangle|\sim 0.4-0.5$, 
(see Fig.~\ref{fig:phase}(c)), indicating a semi-classical nature of magnetism, 
although their spin moments shrink due to quantum fluctuations.
The stripe and vortex phases have the coplanar spins,
and the double-$\bm{Q}$ spiral phase has noncoplanar spins and can be identified as an antiferromagnetic skyrmion lattice phase characterized by the topological charge~\cite{minar2013prb,kathyat2020prb,mukherjee2021prb,kathyat2021prb,mukherjee2022prb}.
\par
In Secs.\ref{sec:weak-coupling}-\ref{sec:dmet}, we employ several analytical and numerical methods, 
which altogether verify our phase diagram; 
we perform the RPA in Sec.~\ref{sec:weak-coupling} and Luttinger-Tisza method in Sec.~\ref{sec:strong-coupling}, respectively, 
which are the reliable approximations 
in the small-$U$ paramagnetic phase and in the large-$U$ limit, respectively. 
We find that both agree well with the phase diagram obtained by the SSDMF. 
On top of that, in Sec.~\ref{sec:dmet} we test the stability of the most delicate phase, SDWI, 
which was not captured in any of the previous works. 
There, we apply the DMET, which is the method that treats nearly full correlation effect 
and can be applied to the incommensurate phase. 
The relationships of these approximations are summarized in 
Sec.~\ref{sec:discussion}, Fig.~\ref{fig:reschart}.
%%%%%%%%%%%%%%%%%%%%%%%%%%%%%
\par
We briefly mention that the periods and types of magnetic phases in our phase diagram 
contradict in several aspects with the previous works based on cluster dynamical mean-field theory (DMFT)~\cite{zhang2015njp} 
and mean field calculation~\cite{minar2013prb}; 
they predict the AFM at small $\theta$ instead of SDWI, and the QSL at around $\theta=\pi$. 
However, we conclude that these phases are the numerical artifacts based on the two findings. 
First, we show in Sec.~\ref{sec:ssdmf} that the calculation using the finite cluster with periodic boundary 
can hardly capture the SDW states with large-scale periods (see also Figs.~\ref{fig:ssdmf}(b)-(d)). 
Since cluster-DMFT treats only the short-range correlation in principle, 
they cannot capture the long-period SDW phases which do not fit to the cluster shape, which is well known. 
Secondly, the QSL phase is concluded to be absent in previous large-scale QMC calculations on the 
$\pi$-flux Hubbard model which is exactly transformed to our model by the local gauge transformation. 
These comparisons are discussed in more detail in Sec.~\ref{sec:discussion}. 
%*%*%*%*%*%*
\subsection{Metal-to-magnetic phase transitions}
The spin splitting of energy bands over the whole Brillouin zone 
differentiates the antisymmetric SOC systems from the systems with inversion symmetry. 
Now we highlight the origin of the above-mentioned various phases as a feature particular to the antisymmetric SOC systems. 
Theoretical details on how we reach these conclusions are shown in Secs.\ref{sec:weak-coupling}-\ref{sec:wilson}.
\par
As we saw in Fig.~\ref{fig:model}(b), there are four Dirac points at 
$(0,0), (0,\pi), (\pi,0), (\pi,\pi)$ which are the $\bm{k}$-points invariant
%under the combination of the time reversal and inversion symmetry operations. 
under the time reversal symmetry operations. 
Two of them lie at the Fermi level at half-filling and stay there throughout the variation of $\theta=[0:\pi]$. 
The other two are rather off from the Fermi level, 
whereas they approach the Fermi level when further increasing $\theta$ and 
finally, they all fall at the Fermi level at $\theta=\pi$, namely $t=0$ and $\lambda=1$. 
\par
The Fermi surfaces carry their spin moments, which appear in pairs for the upper and lower bands crossing 
the Fermi level (not shown here), and along with the variation of the height of the Fermi level at $(0,0)$ and $(\pi,\pi)$ 
with increasing $\theta$, the shape of the major part of the Fermi surface varies 
from square-like ones to smaller pockets. 
At small $\theta$, the nesting instability is the origin of the SDW phases as has been discussed in 
many transition metals including high-$T_c$ cuprates~\cite{bednorz1986zpb,wu2011nature}. 
However, in the present case, the nesting takes place between the Fermi surfaces that carry 
antiparallel spin moments via the incommensurate nesting vector. 
It overwhelms instability to the commensurate AFM ones. 
This {\it spin-pairwise nesting instability} distinguishes the present antisymmetric SOC systems. 
The details will be discussed based on the RPA framework in Sec.~\ref{sec:rpa}.
\par 
The magnetic phase changes its character when $\theta\gtrsim 2\pi/3$. 
Although at the very vicinity of the phase transition, the instability toward the SDWIII phase survives up to $\theta=\pi$, 
it almost simultaneously transforms to the other commensurate stripe or vortex phases. 
This is because, the Fermi pockets become very small and the $(0,0)$ and $(\pi,\pi)$ 
Dirac points become close to the Fermi level. 
In such a situation, the small density of states near the Dirac points and the small Fermi pockets 
no longer has advantages in the Fermi surface nesting instability. 
Then, it is energetically favorable to ``nest" the whole Dirac points of the same shape and 
with this small energy difference to open a band gap. 
This is another remarkable feature of the present system 
which will be disclosed in Sec.~\ref{sec:dirac}.
\par
The magnetic structures and the phase boundaries do not change much with increasing $U$. 
This means that the magnetic properties are already determined by the features of the spin-split bands.
We, however, mention that the energy differences between different magnetic phases are subtle, 
and there are intense competitions between at least two different phases next to each other. 
In the strong coupling limit, the Hubbard Hamiltonian is reduced to the spin Hamiltonian 
with Heisenberg interaction (vary from antiferromagnetic to ferromagnetic ones when $\theta=0$ to $\pi$), 
DM interaction, and bond-dependent AFM Ising-type exchange interactions 
(see Sec.~\ref{sec:spinham}).
Although we deal with them by a semi-classical approximation, the Luttinger-Tisza method,
the ground state of this spin model is in good agreement with the phase diagram, favoring incommensurate (modulated spins) to 
semi-classical spiral and vortex spins 
when moving from small to large $\theta$ (see Sec.~\ref{sec:ltmethod}, Sec.~\ref{sec:mf}, and Sec.~\ref{sec:order}). 

%%%%%%%%%%%%%%%%%%%%%%%%%%%%%%%%%%%%%%%
%%%%%%%%%%%%%%% section %%%%%%%%%%%%%%%
%%%%%%%%%%%%%%%%%%%%%%%%%%%%%%%%%%%%%%%
\section{Weak-coupling approach}
\label{sec:weak-coupling}
%%%%%%%%%% Subsection %%%%%%%%%%
\subsection{Energy band and Fermi surface}
\begin{figure*}[t]
\includegraphics[width=180mm]{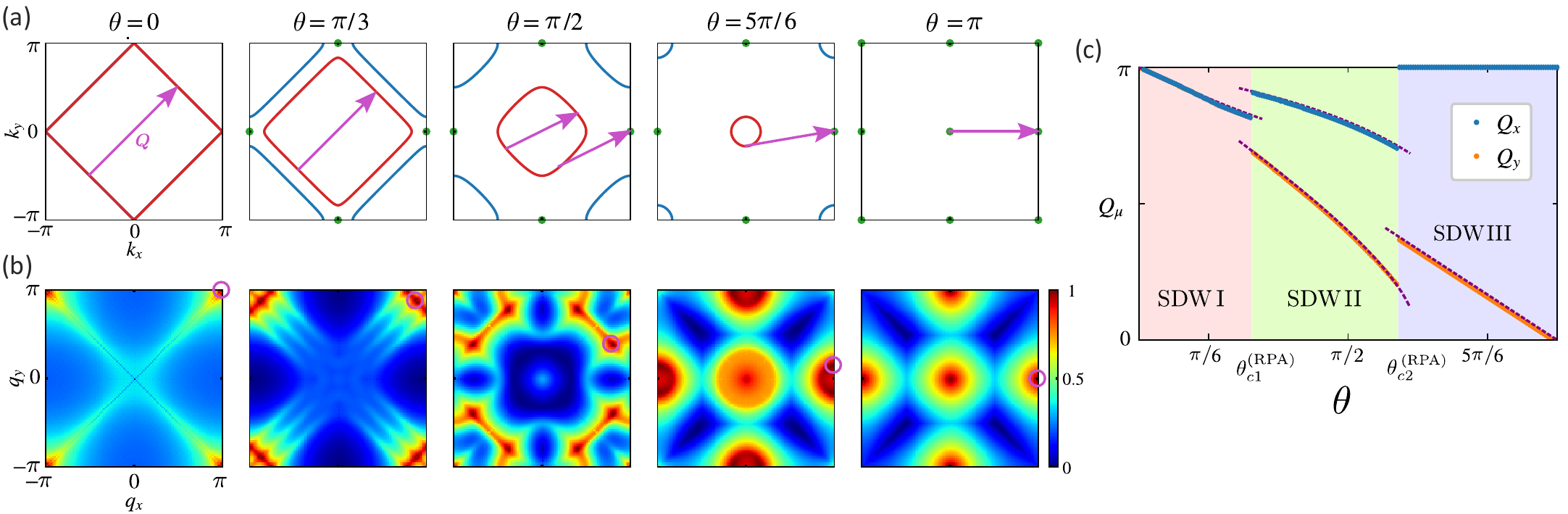}
\caption{(a) Fermi surfaces 
and (b) the density plots of the largest eigenvalues of the bare magnetic susceptibility 
$\lambda_{3}(\bm{q})$ as functions of $\bm{q}$ for several choices of $\theta$.
The purple arrows in the upper panels represent the nesting vector $\bm{Q}$,
and the purple circles in the lower panels represent the peak position of $\lambda_{3}(\bm{q})$ 
within $q_{x}\geq q_{y}\geq0$.
For $\theta\neq0$,
there are two zero-energy Dirac points at $(\pi,0)$ and $(0,\pi)$, 
and other two Dirac points reach the zero-energy 
at $(0,0)$ and $(\pi,\pi)$ for $\theta=\pi$, represented by the green circles. 
The density plot of $\lambda_{3}(\bm{q})$ is normalized to 
$\min_{\bm{q}}\lambda_{3}(\bm{q})=0$ and $\max_{\bm{q}}\lambda_{3}(\bm{q})=1$ 
for clarification. 
(c) Peak position of $\lambda_{3}(\bm{q})$ in the region $q_{x}\geq q_{y}\geq0$ as a function of $\theta$.
There are three phases separated by the boundaries
at $\theta_{c1}^{(\mathrm{RPA})}\sim\pi/4$ and $\theta_{c2}^{(\mathrm{RPA})}\sim2\pi/3$.
The purple dashed lines are the wave numbers 
from Eq.~(\ref{eq:QsdwI}),~(\ref{eq:QsdwII}), and~(\ref{eq:QsdwIII}),
which are in good agreement with those obtained by the RPA.
}
\label{fig:fs}
\end{figure*}

In the weak coupling theory, the non-interacting band structures and the shape of Fermi surfaces 
play the key roles in the metal-to-magnetic phase transitions. 
The $\bm{k}$-space representation of our non-interacting Hamiltonian is
%%%%%%%%%%%%%%%
\begin{align}
\hat{\mathcal{H}}_{0}
=
\sum_{\bm{k}}
\hat{\bm{c}}_{\bm{k}}^{\dagger}
H(\bm{k})
\hat{\bm{c}}_{\bm{k}}
,
\end{align}
%%%%%%%%%%%%%%%
where $\hat{\bm{c}}_{\bm{k}}=(\hat{c}_{\bm{k},\uparrow},\hat{c}_{\bm{k},\downarrow})^{T}$,
$\hat{c}_{\bm{k},\sigma}=(1/\sqrt{N})\sum_{\bm{r}}\hat{c}_{\bm{r},\sigma}\mathrm{e}^{-i\bm{k}\cdot\bm{r}}$,
%%$\bm{k}$ is the two-dimensional wave vector in the Brillouin zone,
and $H(\bm{k})$ is the $2\times2$ Hermitian matrix describing the single-particle Hamiltonian,
%%%%%%%%%%%%%%%
\begin{align}
H(\bm{k})
&=
-2t_{\mathrm{eff}}
\cos(\theta/2)
(
\cos k_{x}
+
\cos k_{y}
)
\sigma^{0}
\nonumber \\
&\hspace{20pt}
-2t_{\mathrm{eff}}
\sin(\theta/2)
(
\sigma^{y}
\sin k_{x}
-
\sigma^{x}
\sin k_{y}
)
.
\end{align}
%%%%%%%%%%%%%%%
The cosine-term is the standard spin-degenerate one and 
the sine-term that originates from the Rashba SOC 
splits the energy band, which is regarded as the $\bm{k}$-dependent Zeeman-term. 
The two energy eigenvalues are
%%%%%%%%%%%%%%%
\begin{align}
\varepsilon_{\pm}(\bm{k})
&=
-2t_{\mathrm{eff}}
\cos(\theta/2)
(
\cos k_{x}
+
\cos k_{y}
)
\nonumber \\
&\hspace{20pt}
\pm
2t_{\mathrm{eff}}
\sin(\theta/2)
\sqrt
{
\sin^{2} k_{x}
+
\sin^{2} k_{y}
}
.
\end{align}
%%%%%%%%%%%%%%%
Figure~\ref{fig:fs}(a) shows the Fermi surfaces for various $\theta$.
The energy bands at $\theta=0$ (without Rashba SOC) are two-fold degenerate 
as a consequence of time reversal and inversion symmetries. 
Finite Rashba SOC ($\theta\neq0$) breaks the inversion symmetry 
and the degenerate bands split except at time-reversal invariant momenta (TRIM)
$\bm{k}_{\mathrm{TRIM}}=(0,0)$, $(\pi,0)$, $(0,\pi)$, and $(\pi,\pi)$,
where $\pm \bm{k}_{\mathrm{TRIM}}$ are the same wave vectors within the first Brillouin zone.
\par
These Kramers degeneracies at the TRIM are protected by the time-reversal symmetry 
and form Dirac points at $\theta\neq0$. 
One can explicitly write down the Dirac Hamiltonian around the TRIM $\bm{k}_{\mathrm{TRIM}}$ as
%%%%%%%%%%%%%%%
\begin{align}
H(\bm{k})
\sim
\varepsilon_{0}(\bm{k}_{\mathrm{TRIM}})
-
2t_{\mathrm{eff}}
\sin\theta
\{
\sigma^{y}
\Delta k_{x}
-
\sigma^{x}
\Delta k_{y}
\}
,
\label{eq:disp}
\end{align}
%%%%%%%%%%%%%%%
where $\varepsilon_{0}(\bm{k})=-2t_{\mathrm{eff}}\cos\theta(\cos k_{x}+\cos k_{y})$
and $\Delta \bm{k}=\bm{k}-\bm{k}_{\mathrm{TRIM}}$. 
Importantly, the velocities are all the same for the four Dirac points. 
Among the four Dirac points,
those at $\bm{k}=(\pi,0)$ and $(0,\pi)$ locate at zero energy level, 
$\varepsilon_{0}(\pi,0)=\varepsilon_{0}(0,\pi)=0$, irrespective of $\theta$. 
These two are protected by the symmetry represented by an antiunitary operator,
%%%%%%%%%%%%%%%
\begin{align}
\hat{\mathcal{T}}
&=
\bigotimes_{\bm{r}}
\left(
\bigotimes_{\sigma}
(
\hat{c}_{\bm{r},\sigma}
-
\mathrm{e}^{i\bm{Q}_{\pi}\cdot\bm{r}}
\hat{c}_{\bm{r},\sigma}^{\dagger}
)
\right)
\exp
\left(
-i\pi
\frac{\hat{S}_{\bm{r}}^{x}+\hat{S}_{\bm{r}}^{y}}{\sqrt{2}}
\right)
K
,
\end{align}
%%%%%%%%%%%%%%%
where $\bm{Q}_{\pi}=(\pi,\pi)$
and $K$ is the complex-conjugation operator, $KiK^{-1}=-i$.
This transformation can be interpreted as the combination
of the time-reversal, particle-hole, and mirror symmetry operations.
The symmetry constraints $[\hat{\mathcal{H}},\hat{\mathcal{T}}]=0$
and the Kramers degeneracy leads to $\varepsilon_{n}(\bm{k})=0$ at these points.
The other two Dirac points have finite energy as
$\varepsilon_{0}(0,0)=-2t_{\mathrm{eff}}\cos(\theta/2)$ and $\varepsilon_{0}(\pi,\pi)=2t_{\mathrm{eff}}\cos(\theta/2)$.
The Fermi pockets related to these two points are centered at $(0,0)$ and $(\pi,\pi)$, 
and as $\theta$ increases, these pockets shrink and 
eventually, the Dirac point reaches the Fermi level at $\theta=\pi$ (strong Rashba SOC limit).
%%%%%%%%%%%%%%%%%%%%%%%%%%%%%%%%%%%%%%%%%%%%%%%%%%%%%%%%%%%%%%%%%%%%%%%%%%%%%%%%%%%%%%%%%%%%%%%%%%%%%%%%%%%%%%%%%%%%%%%%
%%%%%%%%%%%%%%%%%%%%%%%%%%%%%%%%%%%%%%%%%%%%%%%%%%%%%%%%%%%%%%%%%%%%%%%%%%%%%%%%%%%%%%%%%%%%%%%%%%%%%%%%%%%%%%%%%%%%%%%%

%%%%%%%%%% Subsection %%%%%%%%%%
\subsection{Random phase approximation}
\label{sec:rpa}
%%%%% Subsubsection %%%%%
\subsubsection{Magnetic susceptibility}
In the RPA framework, the phase transition is captured by the instability toward the magnetic ordering 
when the magnetic susceptibility $\chi_{\mathrm{RPA}}(\bm{q})$ obtained by the RPA diverges. 
The ordering wave vector $\bm q=\bm Q$ at the critical point that contributes to this divergence 
is usually determined by the shape of the Fermi surface. 
In the standard RPA approach, the energy bands in the metallic phase are spin-degenerate 
and so as the bare magnetic susceptibility, $\chi_0(\bm{q})$. 
The divergence occurs for the nesting vector $\bm Q$ that interpolates the two separate parts of the Fermi surfaces 
that maximally contribute to $\chi_0(\bm{q})$, and 
the magnetic moment is given within the linear response theory as 
$\braket{\hat{\bm{S}}_{\bm{Q}}}= \chi_{\mathrm{RPA}}(\bm{Q})\bm{h}_{\bm{Q}}$, 
with $\chi_{\mathrm{RPA}}(\bm{Q})\propto  (1-2U\chi_0(\bm{Q}))^{-1}$, 
where $\bm{h}_{\bm{Q}}$ is the infinitesimal external magnetic field with the wave vector $\bm{Q}$.
\par
In our case, however, the energy bands carry spin moments that point in various directions depending on $\bm{k}$.
Accordingly, $\chi_{\mathrm{RPA}}(\bm{q})$ and $\chi_{0}(\bm{q})$ are 
no longer scalar but take the form of $3\times 3$ matrices; 
since the system does not have an SU(2) spin-rotational symmetry, 
the $x,y,z$-elements of spin moments are independently taken into account. 
They are given as~\cite{minar2013prb}
%%%%%%%%%%%%%%%
\begin{align}
\chi_{\mathrm{RPA}}(\bm{q})
=
[I_{3\times3}-2U\chi_{0}(\bm{q})]^{-1}\chi_{0}(\bm{q})
,
\end{align}
%%%%%%%%%%%%%%%
with $I_{3\times3}$ being the $3\times3$ identity matrix, 
and 
%%%%%%%%%%%%%%%
\begin{align}
&
\chi_{0}^{\mu\nu}(\bm{q})
\nonumber \\
&=
-
\frac{1}{N}
\hspace{-2pt}
\sum_{\bm{k},n,m}
\hspace{-1pt}
s_{n,m}^{\mu}(\bm{k},\bm{k}+\bm{q})
s_{m,n}^{\nu}(\bm{k}+\bm{q},\bm{k})
F_{n,m}(\bm{k},\bm{k}+\bm{q})
,
\label{eq:chi0}
\end{align}
%%%%%%%%%%%%%%%
for $\mu,\nu=x,y,z$ with 
%%%%%%%%%%%%%%%
\begin{align}
F_{n,m}(\bm{k},\bm{k}+\bm{q})
=
\frac
{f(\varepsilon_{n}(\bm{k}))-f(\varepsilon_{m}(\bm{k}+\bm{q}))}
{\varepsilon_{n}(\bm{k})-\varepsilon_{m}(\bm{k}+\bm{q})+i\hbar\delta}
,
\label{eq:fmn}
\end{align}
%%%%%%%%%%%%%%%
%%%%%%%%%%%%%%%
\begin{align}
s_{n,m}^{\mu}(\bm{k}_{1},\bm{k}_{2})
=
\bm{u}_{n}^{\dagger}(\bm{k}_{1})
\left(
\frac{\sigma^{\mu}}{2}
\right)
\bm{u}_{m}(\bm{k}_{2}). 
\end{align}
%%%%%%%%%%%%%%%
Here $f(\varepsilon)=1/(\exp(\varepsilon/k_{\mathrm{B}}T)+1)$ is the Fermi distribution function 
at temperature $T$, 
$\delta$ is the infinitesimal positive number that represents an adiabatic application of the external field,
and $\bm{u}_{n}(\bm{k})$ is the eigenvector of $H(\bm{k})$ for $\varepsilon_{n}(\bm{k})$ ($n=\pm$) 
given as
%%%%%%%%%%%%%%%
\begin{align}
\bm{u}_{\pm}(\bm{k})
=
\frac{1}{\sqrt{2}}
\begin{pmatrix}
\sqrt{\sin^{2}k_{x}+\sin^{2}k_{y}}\\
\pm(\sin k_{y}-i\sin k_{x})
\end{pmatrix}
.
\end{align}
%%%%%%%%%%%%%%%
\par
We need to find the wave vector $\bm{q}=\bm{Q}$ at which $\chi_{\mathrm{RPA}}(\bm{q})$ diverges.
Since $\chi_{0}(\bm{q})$ is Hermitian, 
there exists a unitary matrix $W(\bm{q})$ that diagonalizes $\chi_{\mathrm{RPA}}(\bm{q})$ as
%%%%%%%%%%%%%%%
\begin{align}
&
W^{\dagger}(\bm{q})
\chi_{\mathrm{RPA}}(\bm{q})
W(\bm{q})
\nonumber \\
&=
\mathrm{diag}
\left(
\frac{\lambda_{1}(\bm{q})}{1-2U\lambda_{1}(\bm{q})},
\frac{\lambda_{2}(\bm{q})}{1-2U\lambda_{2}(\bm{q})},
\frac{\lambda_{3}(\bm{q})}{1-2U\lambda_{3}(\bm{q})}
\right)
,
\end{align}
%%%%%%%%%%%%%%%
where $\lambda_{\ell}(\bm{q})$ is the eigenvalue of $\chi_{0}(\bm{q})$ in an ascending order,
$\lambda_{1}(\bm{q})\leq\lambda_{2}(\bm{q})\leq\lambda_{3}(\bm{q})$.
When we move on to this ``rotating frame'' by $W(\bm{q})$,  the RPA susceptibility becomes diagonal. 
The first diagonal component to diverge in increasing $U$ from $0$ is given by 
$\lambda_{3}(\bm{Q})/(1-2U\lambda_{3}(\bm{Q}))$,
where $\bm{Q}$ is defined as the wave vector that maximizes $\lambda_{3}(\bm{q})$ as
%%%%%%%%%%%%%%%
\begin{align}
\bm{Q}
=
\max_{\bm{q}}
\lambda_{3}(\bm{q})
.
\end{align}
%%%%%%%%%%%%%%%
The Fourier component of the ordered magnetic moment is given by
$\braket{W^{\dagger}(\bm{Q})\hat{\bm{S}}_{\bm{Q}}}\propto(0,0,1)^{T}$ in the ``rotating frame''.
In the original frame,
we have
%%%%%%%%%%%%%%%
\begin{align}
\braket{\hat{\bm{S}}_{\bm{Q}}}
\propto
(
W^{xz}(\bm{Q}),
W^{yz}(\bm{Q}),
W^{zz}(\bm{Q})
)^{T}
,
\end{align}
%%%%%%%%%%%%%%%
and the inverse Fourier transformation yields the magnetic moments in real space.
In the following, we obtain the largest eigenvalue and corresponding eigenvector of the bare magnetic susceptibility $\chi_{0}(\bm{q})$ to determine the properties of the magnetically ordered phase at the critical point. 

%%%%% Subsubsection %%%%%
\subsubsection{Spin-pairwise nesting instability} 
Figure~\ref{fig:fs}(b) shows the maximum eigenvalue of the bare magnetic susceptibility, $\lambda_{3}(\bm{q})$,
as the function of $\bm{q}$ at $k_{B}T=0.05$. 
Here we perform the RPA at finite temperature,
where the temperature simply acts as the smoothing of the step function.
We replace the summation of $\bm{k}$ in Eq.~(\ref{eq:chi0}) by the integral and
evaluated it using the Simpson method. 
The peak position $\bm{Q}$ of $\lambda_{3}(\bm{q})$ shifts as $\theta$ increases, 
which is shown in Fig.~\ref{fig:fs}(c) for $q_{x}\geq q_{y}\geq0$. 
We find three regions separated by the jump of the peak positions, indicating the first order transitions. 
Let us denote the two boundaries as $\theta=\theta_{c1}^{(\mathrm{RPA})}$ and $\theta_{c2}^{(\mathrm{RPA})}$. 
In the SDWI phase at $0\leq\theta\leq\theta_{c1}^{(\mathrm{RPA})}$,
we find the peak along the $q_{x}=q_{y}$ line (see Fig.~\ref{fig:fs}(a) and Fig.~\ref{fig:fs}(b)).
When $\theta=0$, the commensurate AFM order exists at $U>0$ with $\bm{q}=\bm{Q}_{\pi}$. 
As $\theta$ increases, the peak splits and 
the center peak gradually shifts to $q_{x}=q_{y} <\pi$,
indicating an incommensurate magnetic order. 
In the SDWII phase at $\theta_{c1}^{(\mathrm{RPA})}\leq\theta\leq\theta_{c2}^{(\mathrm{RPA})}$. 
the peak position falls off the $q_{x}=q_{y}$ line. 
When the system enters the SDWIII phase at $\theta_{c2}^{(\mathrm{RPA})}\leq\theta<\pi$, 
$\bm Q$ locates at the boundary of the Brillouin zone, 
namely either $Q_x$ or $Q_y$ is equal to $\pi$, 
and finally at $\theta=\pi$ we find the peaks at $(0,\pi)$ and $(\pi,0)$ whose combination forms a vortex ordering. 
\par
These results are consistent with the development of the Fermi surface. 
In the SDWI phase, we can extract the possible nesting vector indicated by arrows in Fig.~\ref{fig:fs}(a).
We denote it as $\bm{Q}_{\mathrm{SDWI}}=(Q_{\mathrm{SDWI}},Q_{\mathrm{SDWI}})$,
which should satisfy $\varepsilon_{+}(\bm{Q}_{\mathrm{SDWI}}/2)=\varepsilon_{+}(-\bm{Q}_{\mathrm{SDWI}}/2)=0$. 
We find the explicit form of $Q_{\mathrm{SDWI}}$ as
%%%%%%%%%%%%%%%
\begin{align}
Q_{\mathrm{SDWI}}
=
2\mathrm{arctan}
\left(
\frac{\sqrt{2}}{\tan(\theta/2)}
\right)
, 
\label{eq:QsdwI}
\end{align}
%%%%%%%%%%%%%%%
which agrees with the value of $\bm Q$ shown in Fig.~\ref{fig:fs}(c) 
at $0\leq\theta\leq\theta_{c1}^{(\mathrm{RPA})}$. 
%%%%%%%%%%%%%%%%%%%%%%%%%%%%%%%%%%%%%%%%%%%%%%%%%%%%%%%%%%%%%%%%%%%%%%%%%%%%%%%%%%%%%%%%%%%%%%%%%%%%%%%%%%%%%%%%%%%%%%%%
\begin{figure}[t]
\includegraphics[width=85mm]{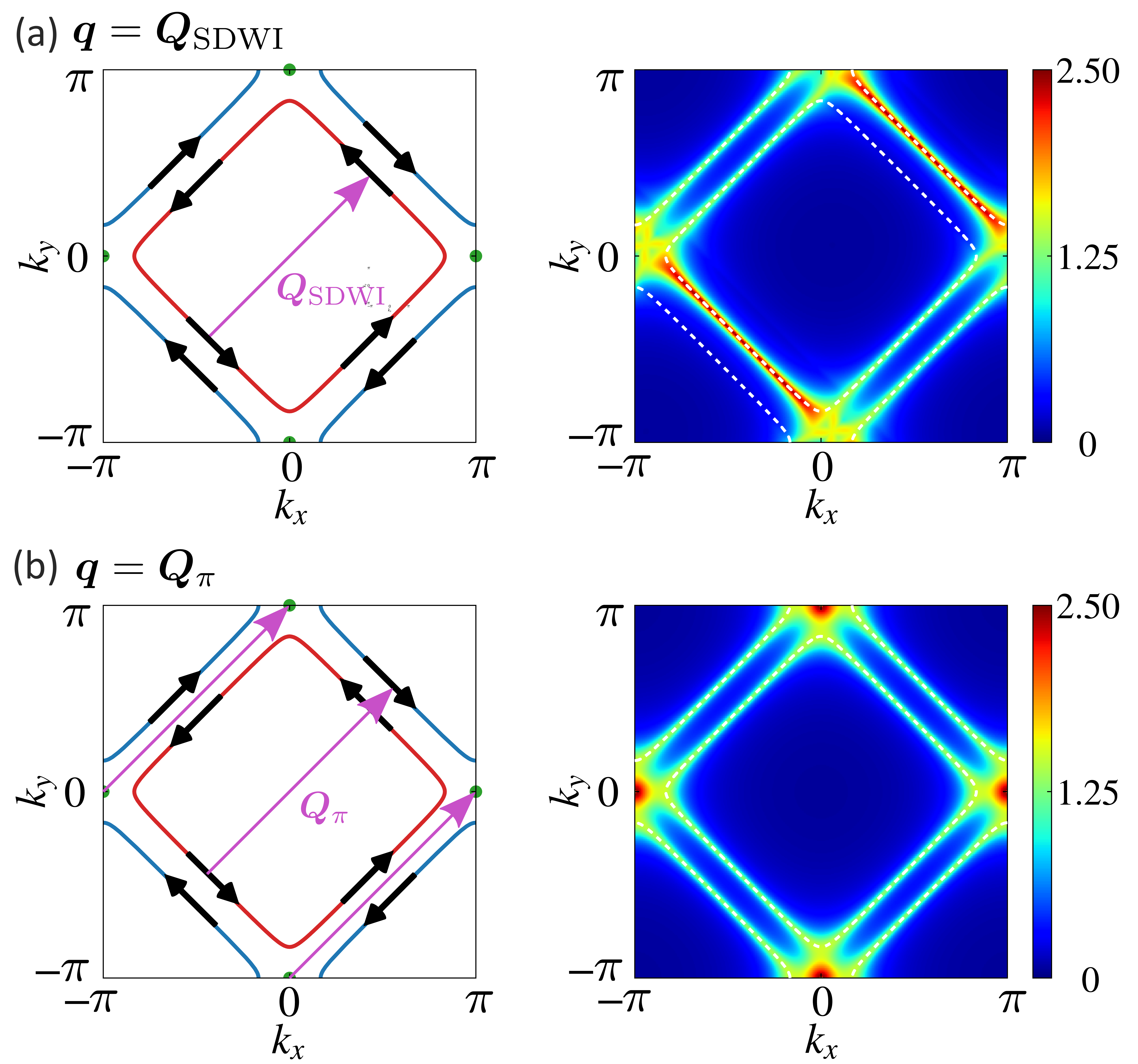}
\caption{Comparison of the two vectors (a) $\bm{Q}_{\mathrm{SDWI}}$ and 
(b) $\bm{Q}_{\pi}$, regarding the contribution to $\chi_0(\bm{Q})$. 
The left panels show the spin orientation (black arrows) on the Fermi surfaces 
and how the nesting vectors connect the Fermi surfaces. 
The right panels show the density plot of the contributions 
to $\lambda_{3}(\bm{Q})$ at $\bm{k}$ which interpolates 
to $\bm{k}+\bm{Q}$ in Eq.(\ref{eq:chi0}). 
Broken line in the right panels is the location of the Fermi surface for the guide to the eye. 
}
\label{fig:pairwise}
\end{figure}
%%%%%%%%%%%%%%%%%%%%%%%%%%%%%%%%%%%%%%%%%%%%%%%%%%%%%%%%%%%%%%%%%%%%%%%%%%%%%%%%%%%%%%%%%%%%%%%%%%%%%%%%%%%%%%%%%%%%%%%%
\par
So far, we have not explicitly discussed the relevance of the spin index of energy bands and the nesting vector. 
However, the spin orientation crucially influences the choice of $\bm{Q}$; 
in Eq.(\ref{eq:chi0}) we find that the nesting from $\bm k$ to $\bm k+\bm q$ 
accompanies the flipping of spins via $\sigma^{\mu}$. 
Figure~\ref{fig:pairwise} shows the contribution of $\bm{Q}_{\mathrm{SDWI}}$ and $\bm{Q}_{\pi}$ to $\lambda_{3}(\bm{q})$;
the states at $\bm k$ combined with $\bm k+\bm q$ contribute to
$F_{n,m}(\bm{k},\bm{k}+\bm{q})$ and $s_{n,m}^{\mu}(\bm{k},\bm{k}+\bm{q})$ in Eq.(\ref{eq:chi0}). 
We evaluate the maximum eigenvalue of the integrand in Eq.(\ref{eq:chi0}) for 
a given $\bm q=\bm{Q}_{\mathrm{SDWI}}$ or $\bm{Q}_{\pi}$ 
as the contribution to $\lambda_{3}(\bm{q})$ at $\bm{k}$. 
The results are shown as the density plots in the right panel.
As shown in the left panel, $\bm{Q}_{\mathrm{SDWI}}$ nests $\bm k$ to $\bm k+\bm q$ belonging to the same Fermi surface 
but carrying opposite spins, 
whereas $\bm{Q}_{\pi}$ nests the two different Fermi surfaces carrying the same spins. 
Apparently, nesting takes place over the edges of the Fermi surface in the former 
but only at an isolated single points $(0,\pi)$ to $(\pi,0)$ for the latter, 
and the former gives larger contributions.
This should be because Eq.(\ref{eq:chi0}) allows off-diagonal (different spin orientations) 
components over a wider range of Fermi surfaces.
Such $\bm{Q}_{\mathrm{SDWI}}$ is generally incommensurate. 
\par
The other two SDW phases are more intriguing. 
The nesting vector in the SDWII phase denoted as $\bm{Q}_{\mathrm{SDWII}}$ 
is expected as the one that crosses $(0,0)$ and connects the two sides 
of the $\varepsilon_+$-Fermi pocket, 
and at the same time, connects one of the points of the Fermi pocket with 
the Dirac point at $(\pi,0)$. 
The condition to be satisfied is, 
%%%%%%%%%%%%%%%
\begin{align}
\varepsilon_{+}(-\bm{Q}_{\mathrm{SDWII}}/2)
&=
\varepsilon_{+}(\bm{Q}_{\mathrm{SDWII}}/2)
\nonumber \\
&=
\varepsilon_{+}((\pi,0)-\bm{Q}_{\mathrm{SDWII}})
=
0
, 
\label{eq:QsdwII}
\end{align}
%%%%%%%%%%%%%%%
which we solved numerically. 
This solution almost perfectly agrees with the peak position in Fig. \ref{fig:fs}(c) 
in the SDWII region. 
\par
The same treatment clarifies the nesting vector $\bm{Q}_{\mathrm{SDWIII}}=(\pi,Q_{\mathrm{SDWIII}})$ 
of the SDWIII at $\theta_{c2}^{(\mathrm{RPA})}\leq\theta$; 
this vector interpolates the Fermi pocket with the Dirac point, 
which should satisfy $\varepsilon_{+}(0,-Q_{\mathrm{SDWIII}})=0$,
and we have 
%%%%%%%%%%%%%%%
\begin{align}
Q_{\mathrm{SDWIII}}
=
\pi-\theta
,
\label{eq:QsdwIII}
\end{align}
%%%%%%%%%%%%%%%
which coincides with the RPA result in Fig.~\ref{fig:fs}(c). 
\par
To summarize, the peak position vector $\bm{Q}$ of $\lambda_3(\bm{q})$ corresponds to the Fermi-surface nesting vector 
in the three SDW phases. 
The one at $\bm{Q}_{\mathrm{SDWI}}$ is a typical nesting vector 
that interpolates the two sides of the square-like Fermi surface. 
However, a crucial difference from the standard SDW is that the nesting occurs between the $\bm k$ and $\bm k+\bm q$ 
points that carry opposite spins, which we call spin-pairwise nesting. 
When we increase $\theta$, these Fermi surfaces shrink and form pockets, 
and the Dirac points come into play.
Accordingly, vector $\bm{Q}_{\mathrm{SDWII}}$ gradually approaches the zone boundary, 
and $\bm{Q}_{\mathrm{SDWIII}}$ with $Q_y=\pi$ becomes very close to the stripe wave vector $(0,\pi)$.
%The reason why the nesting vectors $\bm{Q}_{\mathrm{SDWII}}$ and $\bm{Q}_{\mathrm{SDWIII}}$ 
%favors the zone boundary is that the symmetry-protected Dirac points participate in the instability.  

%%%%% Subsubsection %%%%%
\subsubsection{``Nesting'' of Dirac points}
\label{sec:dirac}
The RPA instability from the metallic phase drives the system to either of the three SDW phases at $0\le \theta <\pi$. 
However, the SDWIII phase at around $5\pi/6 \lesssim \theta <\pi$ is not stable, 
and almost immediately transforms to the stripe or vortex phases in the phase diagram. 
There is an underlying reason to favor these stripe and vortex; 
we saw in Eq.(\ref{eq:disp}) that the four Dirac points have the same velocities 
determined solely by $\theta$. 
Therefore, as the two Dirac points at $(0,0)$ and $(\pi,\pi)$ 
become close to the Fermi level, the vector $\bm Q\sim (0,\pi)$ 
will almost perfectly ``nest" the whole Dirac cone at $(0,0)$ to the one at $(0,\pi)$; 
in Eq.(\ref{eq:chi0}) a substantial range of $\bm k$ near the Dirac points contribute, 
having a small $\varepsilon_{n}(\bm{k})-\varepsilon_{m}(\bm{k}+\bm{q})$ in the denominator of Eq.(\ref{eq:fmn}), 
which will stabilize the stripe-like spin configuration. 
As the pocket shrinks, both $(0,0)$ to $(0,\pi)$ and $(\pi,\pi)$ to $(\pi,0)$ will simultaneously take place. 
This phenomenon is particular to the present system with four Dirac points 
having the same velocities at the special points, protected by the high symmetry. 
The similar energy gain shall work in the SDWII and SDWIII phases slightly off these commensurate 
vectors, which is the reason why the Dirac points which have a very small density of states 
and is very unlikely to participate in the band instability, 
unprecedentedly plays a key role in Fermiology. 

%%%%% Subsubsection %%%%%
\subsubsection{Transverse and Longitudinal susceptibilities}
In the inversion-symmetry-broken systems with SOC, 
the transverse and longitudinal susceptibilities 
in the paramagnetic phase is known to differ because of their spin-split energy bands\cite{Greco2018,Greco2020}. 
Here, we examine how they develop in our case that exhibits the transition to the SDW long-range order. 
The transverse and longitudinal responses to the magnetic field perpendicular/parallel to the $+z$-direction are given as 
\begin{align}
\delta\braket{\hat{S}_{\bm{q}}^{\mu}}
=
\chi^{\perp,\mu}(\bm{q})
\bm{h}_{\bm{q}}^{\mu}
\hspace{5pt}
(\mu=x,y)
,
\hspace{10pt}
\delta\braket{\hat{S}_{\bm{q}}^{z}}
=
\chi^{\parallel}(\bm{q})
\bm{h}_{\bm{q}}^{z}
,
\end{align}
where $\delta\braket{\hat{S}_{\bm{q}}^{\mu}}$ is the deviation of the magnetization from the equilibrium value
and $\bm{h}_{\bm{q}}^{\mu}$ is the magnetic field parallel to the $\mu$-direction ($\mu=x,y,z$).
Here, we focus on the paramagnetic and SDWI phases at small-$\theta$ region. 
For the paramagnetic metallic phase, 
we calculate the susceptibility within the RPA as 
\begin{align}
\chi^{\perp,\mu}(\bm{q})
=
\chi_{\mathrm{RPA}}^{\mu}(\bm{q})
\hspace{5pt}
(\mu=x,y)
,
\hspace{10pt}
\chi^{\parallel}(\bm{q})
=
\chi_{\mathrm{RPA}}^{zz}(\bm{q})
.
\end{align}
These two formulas, however, do not hold in the SDWI phase since the magnetic order breaks 
the translational symmetry of the system. 
In such case, we take the derivative of $\braket{\hat{S}_{\bm{q}}^{\mu}}$ ($\mu=x,y,z$)
with respect to the infinitesimal magnetic field as 
\begin{align}
\chi^{\perp,\mu}(\bm{q})
=
\left.
\frac{\partial\braket{\hat{S}_{\bm{q}}^{\mu}}}{\partial|\bm{h}_{\bm{q}}^{\mu}|}
\right|_{\bm{h}_{\bm{q}}^{\mu}=\bm{0}}
,
\hspace{10pt}
\chi^{\parallel}(\bm{q})
=
\left.
\frac{\partial\braket{\hat{S}_{\bm{q}}^{z}}}{\partial|\bm{h}_{\bm{q}}^{z}|}
\right|_{\bm{h}_{\bm{q}}^{z}=\bm{0}}
\end{align}
\par
\begin{figure}[t]
\includegraphics[width=90mm]{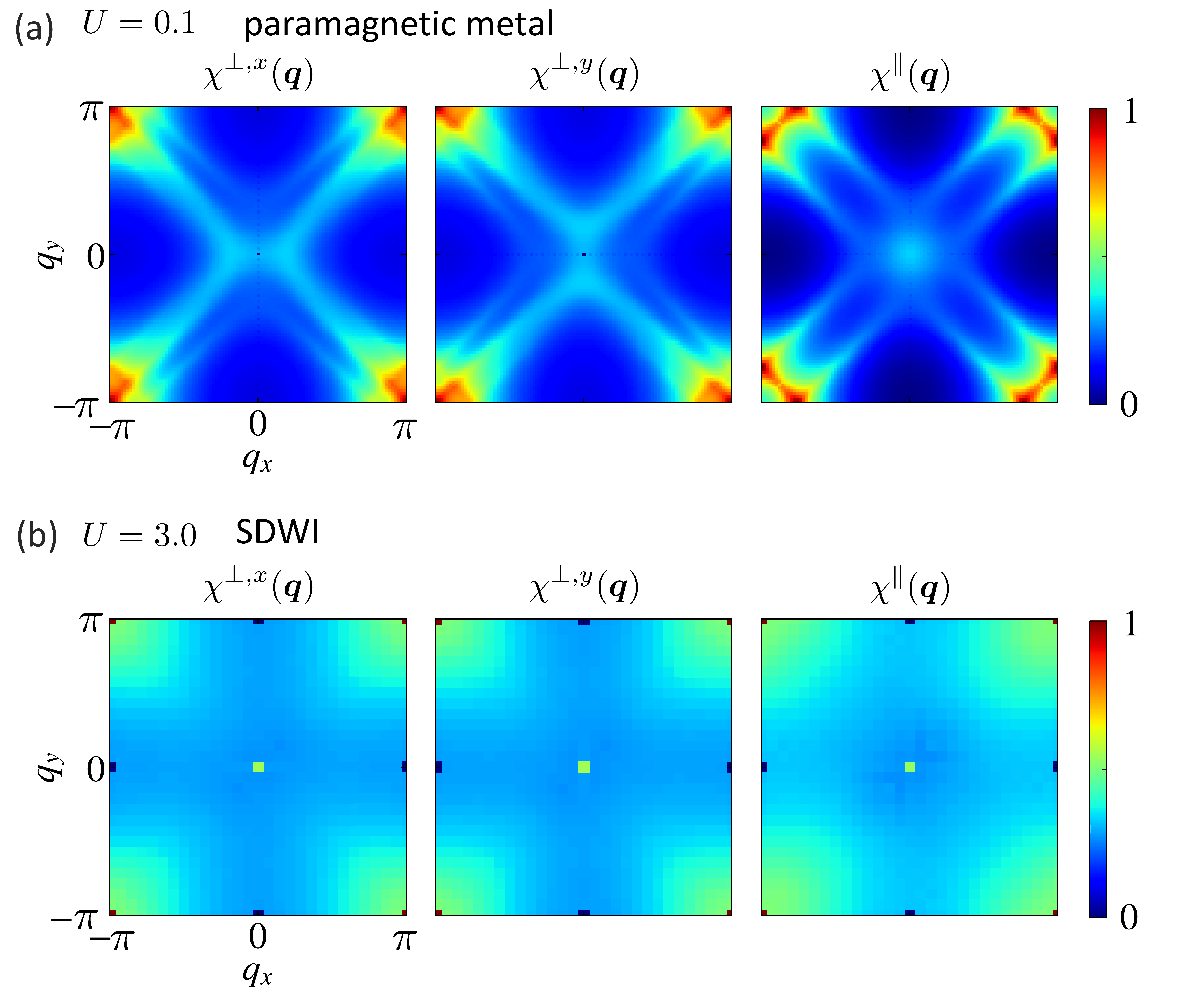}
\caption{
Transverse and longitudinal susceptibilities 
 (a) $U=0.1$ (metal) and (b) $U=3.0$ (SDWI) at $\theta=\pi/6$ and $k_{\mathrm{B}}T=0.05$. 
For the paramagnetic metallic phase,
the peak position locates at $\bm{q}=(\pi,\pi)$ for the transverse susceptibilities,
and $\bm{q}=(\pm\pi,\pm q^{*}),(\pm q^{*},\pm\pi)$ ($q^{*}\neq\pi$) for the longitudinal one.
For the SDWI phase, 
both the transverse and longitudinal susceptibilities have the peak at $\bm{q}=(\pi,\pi)$.
The density plot is normalized for clarification.
}
\label{fig:sus}
\end{figure}

Figures~\ref{fig:sus}(a) and \ref{fig:sus}(b) show 
the transverse and longitudinal susceptibilities at $\theta=\pi/6$ and $k_{\mathrm{B}}T=0.05$ 
with $U=0.1$ and 3 which correspond to the paramagnetic and SDWI phases, respectively. 
In the paramagnetic phase, the transverse susceptibility has the peak at $\bm{q}=(\pi,\pi)$,
whereas the peak position of the longitudinal susceptibility has an incommensurate wave vector. 
This result indicates that the system favors the AFM order in the $xy$-plane, 
but the finite $z$-component of the spin leads to the incommensurate magnetic ordering.
It has relevance to the experimental observation in Ba$_{2}$XGe$_{2}$O$_{7}$ ($X=$Cu,Co,Ge) 
which are the square lattice antiferromagnets with the DM interaction and easy-plane anisotropy
\cite{zheludev2003prb,masuda2010prb,murakawa2012prb}.
In the case of $X=$Co and Ge, the spin are confined in the $xy$-plane by the easy-plane anisotropy 
and show an AFM order at low temperature. 
For $X=$Cu, $S=1/2$ shows no anisotropy and the finite $z$-component of spin moments
leads to an incommensurate spiral order~\cite{zheludev2003prb,masuda2010prb,murakawa2012prb}.
\par
When we enter the SDWI phase, 
both the transverse and longitudinal susceptibilities turn out to have the peak at $\bm{q}=(\pi,\pi)$ 
which do not seem to show much substantial difference with each other except for 
the slight difference in their width. 
The incommensurate peaks in the paramagnetic phases disappear once they order.
The remaining fluctuations off the ordered spin moments develop relatively uniformly for all spatial directions, 
which originate from the antiferromagnetic Heisenberg exchange. 
The previous DMFT result predicted the AFM phase in the same parameter region possibly reflecting this 
secondary important correlation,
which should be because they could not capture the SDWI~\cite{zhang2015njp}.
%%%%%%%%%%%%%%%%%%%%%%%%%%%%%%%%%%%%%%%%%%%%%%%%%%%%%%%%%%%%%%%%%%%%%%%%%%%%%%%%%%%%%%%%%%%%%%%%%%%%%%%%%%%%%%%%%%%%%%%%
%%%%%%%%%%%%%%%%%%%%%%%%%%%%%%%%%%%%%%%%%%%%%%%%%%%%%%%%%%%%%%%%%%%%%%%%%%%%%%%%%%%%%%%%%%%%%%%%%%%%%%%%%%%%%%%%%%%%%%%%

%%%%%%%%%%%%%%%%%%%%%%%%%%%%%%%%%%%%%%%
%%%%%%%%%%%%%%% section %%%%%%%%%%%%%%%
%%%%%%%%%%%%%%%%%%%%%%%%%%%%%%%%%%%%%%%
\section{Strong-coupling approach}
\label{sec:strong-coupling}

%%%%%%%%%% Subsection %%%%%%%%%%
\subsection{Spin Hamiltonian in the strong coupling limit}
\label{sec:spinham}
At $U/t_{\mathrm{eff}} \rightarrow\infty$, 
the Mott insulator with one electron per site is realized. 
The primary interactions between the localized spin moments 
are the kinetic exchange, which has both symmetric and antisymmetric terms. 
The effective spin model is obtained 
by a degenerate perturbation theory~\cite{kaplan1983zpb,shekhtman1993prb}
and we have 
%%%%%%%%%%%%%%%
\begin{align}
\hat{\mathcal{H}}_{\mathrm{spin}}
=
\sum_{\bm{r}}
\sum_{\mu}
\left\{
J
\hat{\bm{S}}_{\bm{r}}
\cdot
\hat{\bm{S}}_{\bm{r}+\bm{e}_{\mu}}
-
\bm{D}_{\mu}
\cdot
(
\hat{\bm{S}}_{\bm{r}}
\times
\hat{\bm{S}}_{\bm{r}+\bm{e}_{\mu}}
)
\right.
\nonumber \\
\left.
+
\left(
\sqrt{J^{2}+D^{2}}-J
\right)
(\bm{n}_{\mu}\cdot\hat{\bm{S}}_{\bm{r}})
(\bm{n}_{\mu}\cdot\hat{\bm{S}}_{\bm{r}+\bm{e}_{\mu}})
\right\}
,
\label{eq:Hspin}
\end{align}
%%
%%%%%%%%%%%%%
where the parameters $J$ and $\bm{D}_{\mu}$ are defined as
%%%%%%%%%%%%%%%
\begin{align}
J
=
\frac{4t_{\mathrm{eff}}^{2}}{U}
\cos\theta
,
\hspace{20pt}
\bm{D}_{\mu}
=
\left(
\frac{4t_{\mathrm{eff}}^{2}}{U}
\sin\theta
\right)
\bm{n}_{\mu}
,
%%\label{}
\end{align}
%%%%%%%%%%%%%%%
and $D=|\bm{D}_{\mu}|=(4t_{\mathrm{eff}}^{2}/U)\sin\theta$.
The effect of Rashba SOC appears as the DM interaction and bond-dependent Ising-type exchange interaction
in the second and third terms of Eq.~(\ref{eq:Hspin}), respectively.
The spin Hamiltonian we dealt as the strong coupling limit of the model
seems consists of the same types of terms with the strong double-exchange limit
of the Kondo lattice model with Rashba SOC~\cite{kathyat2020prb,mukherjee2021prb,kathyat2021prb,mukherjee2022prb},
while the weak-SOC region seems to suffer the same difficulty as the Hubbard model
about dealing with the incommensurate orders in a finite size calculations. 
Figure~\ref{fig:lt}(a) shows the $\theta$-dependence of the three interaction strength,
$J$, $D$, and $\sqrt{J^{2}+D^{2}}-J$.
The Heisenberg exchange $J$ and DM interaction $D$ play a major role at small $\theta$. 
At $\theta>\pi/2$, 
the bond-dependent Ising-type exchange $\sqrt{J^{2}+D^{2}}-J$ becomes dominant, 
and the classical ground state is expected to be the stripe or vortex-like states.
The effective Hamiltonian Eq.(\ref{eq:Hspin}) can be rewritten as
%%%%%%%%%%%%%%%
\begin{align}
\hat{\mathcal{H}}_{\mathrm{spin}}
&=
J_{\mathrm{eff}}
\sum_{\bm{r}}
\left(
\hat{\bm{S}}_{\bm{r}}^{T}
R^{y}(\theta)
\hat{\bm{S}}_{\bm{r}+\bm{e}_{x}}
+
\hat{\bm{S}}_{\bm{r}}^{T}
R^{x}(-\theta)
\hat{\bm{S}}_{\bm{r}+\bm{e}_{y}}
\right)
,
\label{eq:Hspin-R}
\end{align}
%%%%%%%%%%%%%%%
where $J_{\mathrm{eff}}=(4t_{\mathrm{eff}}^{2}/U)$ and $R^{\mu}(\theta)$ are 
the three-dimensional rotation matrix that yields the $\theta$-rotation about the $\mu$-axis ($\mu=x,y,z$).
%%%%%
\subsection{Luttinger-Tisza method}
\label{sec:ltmethod}
We first examine the classical ground-state by the Luttinger-Tisza method~\cite{luttinger1946pr,luttinger1951pr,lyons1960pr}.
Let us approximate the ground state as a product state $\ket{\Phi_{\mathrm{MF}}}=\bigotimes_{\bm{r}}\ket{\bm{m}_{\bm{r}}}$,
where $\ket{\bm{m}_{\bm{r}}}$ is the spin coherent state whose expectation value is given as
$\braket{\bm{m}_{\bm{r}}|\hat{\bm{S}}_{\bm{r}}|\bm{m}_{\bm{r}}}=S\bm{m}_{\bm{r}}$ with $S=1/2$.
The unit vector $\bm{m}_{\bm{r}}$ represents the direction of the classical spin.
The classical ground-state energy
$E_{\mathrm{MF}}=\braket{\bm{m}_{\bm{r}}|\hat{\mathcal{H}}_{\mathrm{spin}}|\bm{m}_{\bm{r}}}$ can be obtained
by replacing the spin operator $\hat{\bm{S}}_{\bm{r}}$ to $S\bm{m}_{\bm{r}}$ in Eq.~(\ref{eq:Hspin-R}). 
In the Luttinger-Tisza method, 
we first minimize the classical ground-state energy under the global constraint $\sum_{\bm{r}}\bm{m}_{\bm{r}}^{2}=N$. 
Then we check whether the obtained solution satisfies the local constraint, $\bm{m}_{\bm{r}}^{2}=1$. 
We introduce the Lagrange multiplier $\lambda$ to $E_{\mathrm{MF}}$ 
and minimize the following function, 
%%%%%%%%%%%%%%%
\begin{align}
E_{\mathrm{LT}}
=
E_{\mathrm{MF}}
-
\lambda
\left(
\sum_{\bm{r}}
|\bm{m}_{\bm{r}}|^{2}
-
N
\right)
.
\end{align}
%%%%%%%%%%%%%%%
Using the Fourier transformation, 
%%%%%%%%%%%%%%%
$\bm{m}_{\bm{r}}
=
\sum_{\bm{q}}
\bm{m}(\bm{q})
\mathrm{e}^{i\bm{q}\cdot\bm{r}}$
,
%%%%%%%%%%%%%%%
it is rewritten as 
%%%%%%%%%%%%%%%
\begin{align}
E_{\mathrm{LT}}
=
\lambda
N
+
N
\sum_{\bm{q}}
\bm{m}^{\dagger}(\bm{q})
\left(
F(\bm{q})
-
\lambda I
\right)
\bm{m}(\bm{q})
,
\end{align}
%%%%%%%%%%%%%%%
with $3\times3$ Hermitian matrix,
%%%%%%%%%%%%%%%
\begin{align}
F(\bm{q})
&=
\frac{J_{\mathrm{eff}}S^{2}}{2}
\left(
R^{y}(\theta)
\mathrm{e}^{iq_{x}}
+
R^{x}(-\theta)
\mathrm{e}^{iq_{y}}
+
\mathrm{h.c.}
\right)
.
\end{align}
%%%%%%%%%%%%%%%
We choose a vector $\bm{q}=\bm{Q}_{\mathrm{LT}}$ that minimizes the lowest eigenvalue of $F(\bm{q})$,
and denote the corresponding eigenvector as $\bm{m}_{0}(\bm{Q}_{\mathrm{LT}})$.
The classical spin configuration that minimizes $E_{\mathrm{LT}}$ can be generally written as
%%%%%%%%%%%%%%%
\begin{align}
\bm{m}_{\bm{r}}
=
\sum_{\bm{Q}_{\mathrm{LT}}}
v(\bm{Q}_{\mathrm{LT}})
\bm{m}_{0}(\bm{Q}_{\mathrm{LT}})
\mathrm{e}^{i\bm{Q}_{\mathrm{LT}}\cdot\bm{r}}
,
\end{align}
%%%%%%%%%%%%%%%
where the coefficient $v(\bm{Q}_{\mathrm{LT}})\in\mathbb{C}$ are determined by imposing the global constraints.
For these solutions, we need to check whether $\bm{m}_{\bm{r}}$ satisfies the local constraint $\bm{m}_{\bm{r}}^{2}=1$. 
%%%%%%%%%%%%%%%%%%%%%%%%%%%%%%%%%%%%%%%%%%%%%%%%%%%%%%%%%%%%%%%%%%%%%%%%%%%%%%%%%%%%%%%%%%%%%%%%%%%%%%%%%%%%%%%%%%%%%%%%
\begin{figure}[t]
\includegraphics[width=85mm]{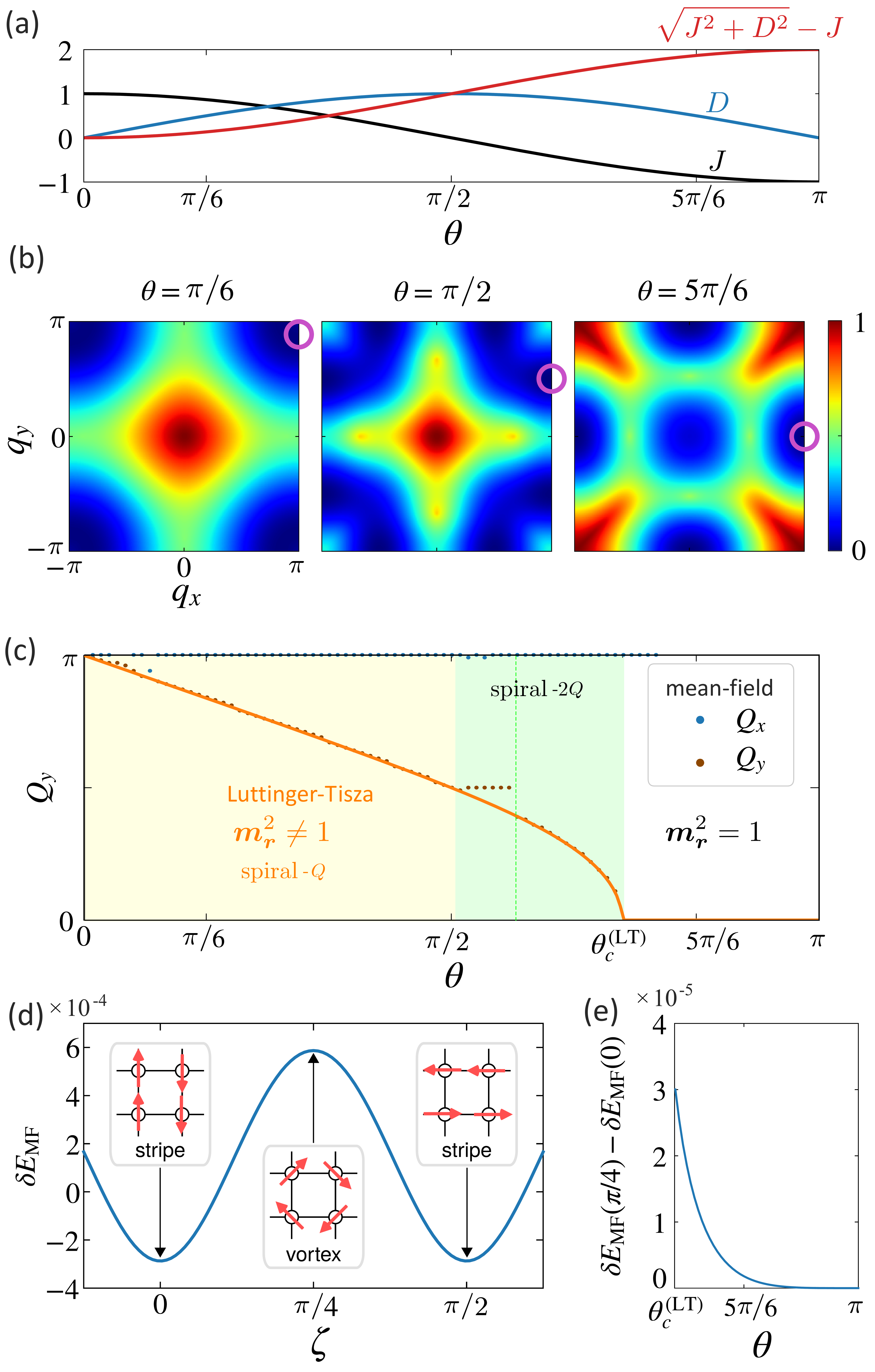}
\caption{
(a) Heisenberg exchange intearction $J$, DM interaction $D$, and bond-dependent Ising-type exchange interaction $\sqrt{J^{2}+D^{2}}-J$ as functions of $\theta$.
(b) Density plot of the lowest eigenvalue of $F(\bm{q})$ as a function of $\bm{q}$ for $\theta=\pi/6$, 
$\theta=\pi/2$, and $\theta=5\pi/6$, with its values normalized to $[0:1]$ 
(c) Ordering wave vector $\bm{Q}_{\mathrm{LT}}$ at $q_{x}\geq q_{y}\geq0$ as a function of $\theta$ 
obtained by the Luttinger-Tisza method (solid line): those in the spiral-$Q$ region 
$\theta<\theta_{c}^{(\mathrm{LT})}$ have $\bm{m_r}<1$. 
Data points are obtained as the mean-field solution, and show the spiral-$2Q$ 
at $\pi/2<\theta<\theta_{c}^{(\mathrm{LT})}$: 
a small plateau and other regions have different $Q$ but they are both the spiral-$2Q$. 
(d) Lowest order quantum energy correction $\delta E_{\mathrm{MF}}$ from the linear spin-wave theory.
(e) Energy difference between the stripe ($\zeta=0$) and vortex ($\zeta=\pi/4$) phases.
}
\label{fig:lt}
\end{figure}
%%%%%%%%%%%%%%%%%%%%%%%%%%%%%%%%%%%%%%%%%%%%%%%%%%%%%%%%%%%%%%%%%%%%%%%%%%%%%%%%%%%%%%%%%%%%%%%%%%%%%%%%%%%%%%%%%%%%%%%%
\par
Figure~\ref{fig:lt}(b) shows the lowest eigenvalue of $F(\bm{q})$ as a function of $\bm{q}$
for $\theta=\pi/6$, $\pi/2$, and $5\pi/6$. 
The minimum value among them is realized at the Brillouin zone boundary marked with purple circles, 
which give $\bm{Q}_{\mathrm{LT}}$. 
The analytical form of this classical ordering wave vector is obtained as 
$\bm{Q}_{\mathrm{LT}}=(\pi,Q_{\mathrm{LT}})$, where 
%%%%%%%%%%%%%%%
\begin{align}
Q_{\mathrm{LT}}
=
\mathrm{arccos}
\left(
-
\frac{\cos\theta}{2\cos(\theta/2)}
\sqrt{1+3\cos^{2}(\theta/2)}
\right)
,
\label{eq:QLT}
\end{align}
%%%%%%%%%%%%%%%
for $0\leq\theta\leq\theta_{c}^{(\mathrm{LT})}$, 
and $Q_{\mathrm{LT}}=0$ for $\theta_{c}^{(\mathrm{LT})}\leq\theta$. 
Here,
$\theta_{c}^{(\mathrm{LT})}$ is the threshold 
at which the argument of the inverse cosine function in Eq.~(\ref{eq:QLT}) exceeds the range $[-1,1]$. 
Figure \ref{fig:lt}(c) shows Eq.(\ref{eq:QLT}) as a function of $\theta$. 
The wave number $Q_{\mathrm{LT}}$ monotonically decreases from $\pi$ as $\theta$ increases,
and becomes $0$ for $\theta>\theta_{c}^{(\mathrm{LT})}$. 
This result indicates that the system shows an incommensurate magnetic order for 
$0<\theta<\theta_{c}^{(\mathrm{LT})}$,
and a commensurate order for $\theta>\theta_{c}^{(\mathrm{LT})}$.
\par
Since $F(\bm{Q}_{\mathrm{LT}})$ is block-diagonal, 
the corresponding eigenvector is obtained by diagonalizing
the $2\times2$ block matrix and we find 
%%%%%%%%%%%%%%%
\begin{align}
\bm{m}_{0}(\bm{Q}_{\mathrm{LT}})
=
\begin{pmatrix}
0\\
\sin\xi(\theta)/2\\
i\cos\xi(\theta)/2
\end{pmatrix}
,
\end{align}
%%%%%%%%%%%%%%%
where $\xi(\theta)$ is defined as
%%%%%%%%%%%%%%%
\begin{align}
\xi(\theta)
%\nonumber \\
=\!
\mathrm{arctan}
\bigg(\!\!
-\!
\frac
{\sqrt{4\cos^{2}(\theta/2)-\cos^{2}\theta(1+3\cos^{2}(\theta/2))}}
{\sin(\theta/2)}
\bigg)
,
\end{align}
%%%%%%%%%%%%%%%
for $0\leq\theta\leq\theta_{c}^{(\mathrm{LT})}$,
and $\xi(\theta)=\pi$ for $\theta_{c}^{(\mathrm{LT})}\leq\theta$. 
The value of $\theta_{c}^{(\mathrm{LT})}= 131.8^\circ$ 
is almost identical to the stripe-vortex phase boundary at $U=8$ in the SSDMF phase diagram. 
\par
Let us examine the spin configuration in the classical ground state. 
%The present formulation is limited to the description of a single-$\bm{Q}$ state.
Here we only consider single-$\bm{Q}$ states,
but one can apply the same arguments to double-$\bm{Q}$ and higher-order-$\bm{Q}$ states.
In the region $0<\theta<\theta_{c}^{(\mathrm{LT})}$, 
the magnetic moment under the global constraint is given by the combination of $\pm \bm{Q}_{\mathrm{LT}}$ 
having $\bm{m}_{\bm{q}}=\bm{m}_{-\bm{q}}^{*}$ as
%%%%%%%%%%%%%%%
\begin{align}
&
\bm{m}_{\bm{r}}=
\nonumber \\
&\sqrt{2}
\cos\pi x
\hspace{-2pt}
\left(
\bm{e}^{y}
\sin\frac{\xi(\theta)}{2}
\cos Q_{\mathrm{LT}}y
-
\bm{e}^{z}
\cos\frac{\xi(\theta)}{2}
\sin Q_{\mathrm{LT}}y
\right)
.
\end{align}
%%%%%%%%%%%%%%%
However, this solution 
has $\bm{m}_{\bm{r}}^{2}=1-\cos\xi(\theta)\cos2Q_{\mathrm{LT}}y\neq 1$
and does not satisfy the local constraint.
To obtain the proper classical ground state satisfying $\bm{m}_{\bm{r}}^{2}=1$,
we need to add the higher-harmonics components~\cite{liu2016prb},
while it is expected that $\bm{Q}_{\mathrm{LT}}$ still represents the dominant wave vector in $\bm{m}_{\bm{r}}$. 
\par
At $\theta>\theta_{c}^{(\mathrm{LT})}$,
the Luttinger-Tisza method suggests a short-period 
magnetic order represented by the wave vector $(\pi,0)$ or $(0,\pi)$. 
These two solutions yielding 
$\bm{m}_{\bm{r}}=\bm{e}^{y}\cos\pi x$ for the former 
and $\bm{m}_{\bm{r}}=-\bm{e}^{x}\cos\pi y$ for the latter 
are degenerate in energy.
Since the system does not have an SU(2) symmetry,
this degeneracy is accidental and is lifted by the quantum fluctuations or by adding higher-order perturbations.
The resultant lowest energy state (if we neglect higher energy states for perturbation) 
takes the form of the linear combination of the two as 
%%%%%%%%%%%%%%%
\begin{align}
\bm{m}_{\bm{r}}
=
\bm{e}^{y}
\cos\pi x
\cos\zeta
-
\bm{e}^{x}
\cos\pi y
\sin\zeta
,
\label{eq:m_lt}
\end{align}
%%%%%%%%%%%%%%%
where $\zeta\in[0,2\pi)$ denotes the relative weights between the two solutions. 

%%%%%%%%%% Subsection %%%%%%%%%%
\subsection{Mean-field approximation for $0\leq\theta\leq\theta_{c}^{(\mathrm{LT})}$}
\label{sec:mf}
We found that the solution of the Luttinger-Tisza method does not fulfill the local constraint 
for the classical ground state at $0\leq\theta\leq\theta_{c}^{(\mathrm{LT})}$. 
It is possibly because the incommensurate spin structure represented by 
$\bm{Q}_{\mathrm{LT}}$ is beyond the classical description. 
Furthermore, it is natural to have nonuniform $|\bm{m}_{\bm{r}}|<1$ when the Hamiltonian (\ref{eq:Hspin-R}) 
is treated quantum mechanically.
At $\pi/2 \lesssim \theta \le \theta_{c}^{(\mathrm{LT})}$, however, there is a spiral-$2Q$ phase in the phase diagram,
which can be described within the classical framework.
Indeed, there are some examples that local constraints recover by simply adding higher-harmonics componenets~\cite{liu2016prb}.
\par
We thus make corrections to the Luttinger-Tisza method by applying a mean-field treatment to 
Eq.(\ref{eq:Hspin-R}), which automatically satisfies the constraint $\bm{m}_{\bm{r}}^{2}=1$. 
The mean-field Hamiltonian is given as
%%%%%%%%%%%%%%%
\begin{align}
\hat{\mathcal{H}}_{\mathrm{MF}}
=
-
\sum_{\bm{r}}
\sum_{\mu}
\braket{\hat{\bm{S}}_{\bm{r}}}^{T}
F_{\mu}
\braket{\hat{\bm{S}}_{\bm{r}+\bm{e}_{\mu}}}
-
\sum_{\bm{r}}
\bm{h}_{\bm{r}}^{(\mathrm{MF})}
\cdot
\hat{\bm{S}}_{\bm{r}}
,
\end{align}
%%%%%%%%%%%%%%%
where $F_{x}=R^{y}(\theta)$ and $F_{y}=R^{x}(-\theta)$. 
The first term is a constant and $\bm{h}_{\bm{r}}^{(\mathrm{MF})}$ is the mean field 
determined by the expectation value of the spin around $\bm{r}$,
%%%%%%%%%%%%%%%
\begin{align}
\bm{h}_{\bm{r}}^{(\mathrm{MF})}
=
-
\sum_{\mu}
\left(
\braket{\hat{\bm{S}}_{\bm{r}-\bm{e}_{\mu}}}^{T}
F_{\mu}
+
\braket{\hat{\bm{S}}_{\bm{r}+\bm{e}_{\mu}}}^{T}
F_{\mu}^{T}
\right)
.
\label{eq:hr_eff_lt}
\end{align}
%%%%%%%%%%%%%%%
We iteratively minimize the energy by evaluating the magnetization self-consistently as 
$\bm{m}_{\bm{r}}=-\bm{h}_{\bm{r}}^{(\mathrm{MF})}/|\bm{h}_{\bm{r}}^{(\mathrm{MF})}|$ at each step.
Figure \ref{fig:lt}(c) shows the ordering wave vector $(Q_x,Q_y)$ as data points.
They agree well with the Luttinger-Tisza line,
and are also consistent with the previous results in the small-$\theta$ region~\cite{farrell2016prb}.
At $\pi/2<\theta < \theta_{c}^{(\mathrm{LT})}$, we newly find a spiral-$2Q$ phase that was not captured in the
Luttinger-Tisza method, which agrees with the phase diagram in the weaker coupling region. 

%%%%%%%%%% Subsection %%%%%%%%%%
\subsection{Order by quantum disorder}
\label{sec:order}
We need to determine the classical ground state at $\theta\geq\theta_{c}^{(\mathrm{LT})}$ 
by fixing the value of $\zeta$ in Eq.(\ref{eq:m_lt}). 
The correction from the degenerate $E_{\mathrm{MF}}$ is evaluated by the 
linear spin-wave theory that accounts for the lowest order quantum fluctuation energy. 
Starting from the magnetically ordered classical state parameterized by $\zeta$ in Eq.(\ref{eq:m_lt}), 
we first rotate the spin quantization axis to the direction of the ordered moment 
by a matrix $R_{\bm{r}}$, and perform 
the Holstein-Primakoff transformation~\cite{holstein1940} in the rotating frame as 
%%%%%%%%%%%%%%%
\begin{align}
&
R_{\bm{r}}
\hat{\bm{S}}_{\bm{r}}
\nonumber \\
&\simeq
\sqrt{\frac{S}{2}}
(
\hat{b}_{\bm{r}}
+
\hat{b}_{\bm{r}}^{\dagger}
)
\bm{e}^{x}
-i
\sqrt{\frac{S}{2}}
(
\hat{b}_{\bm{r}}
-
\hat{b}_{\bm{r}}^{\dagger}
)
\bm{e}^{y}
+
(S-\hat{b}_{\bm{r}}^{\dagger}\hat{b}_{\bm{r}})
\bm{e}^{z}
,
\end{align}
%%%%%%%%%%%%%%%
where the bosonic operator $\hat{b}_{\bm{r}}$ ($\hat{b}_{\bm{r}}^{\dagger}$) represents 
the annihilation (creation) of the magnon at site $\bm{r}$. 
The higher-order terms that contribute to the magnon-magnon interactions~\cite{zhitomirsky2013} are usually 
irrelevant for the present discussion. 
The spin Hamiltonian can be approximated as $\hat{\mathcal{H}}\simeq E_{\mathrm{MF}}+\hat{\mathcal{H}}_{\mathrm{mag.}}$,
where $\hat{\mathcal{H}}_{\mathrm{mag.}}$ is the quadratic form of the bosonic Hamiltonian.
The Bogoliubov transformation of $\hat{b}_{\bm{r}}$
leads to the diagonalized form of $\hat{\mathcal{H}}_{\mathrm{mag.}}$ as~\cite{colpa1978}
%%%%%%%%%%%%%%%
\begin{align}
\hat{\mathcal{H}}_{\mathrm{mag.}}
=
E_{\mathrm{qc1}}
+
\sum_{\bm{k}}
\sum_{\ell=1}^{4}
\varepsilon_{\ell}^{(\mathrm{mag.})}(\bm{k})
\left(
\hat{\gamma}_{\bm{k},\ell}^{\dagger}
\hat{\gamma}_{\bm{k},\ell}
+
\frac{1}{2}
\right)
,
\end{align}
%%%%%%%%%%%%%%%
where $E_{\mathrm{qc1}}$ is the constant term,
$\hat{\gamma}_{\bm{k},\ell}$ is the new bosonic operator that is written as the linear combination of $\hat{b}_{\bm{r}}$
and $\hat{b}_{\bm{r}}^{\dagger}$,
and $\varepsilon_{\ell}^{(\mathrm{mag.})}(\bm{k})$ is the magnon band.
Here, the quantum correction to the classical ground state $\delta E_{\mathrm{MF}}$ 
is given by the constant shift and the zero-point fluctuation as 
%%%%%%%%%%%%%%%
\begin{align}
\delta E_{\mathrm{MF}}
=
E_{\mathrm{qc1}}
+
\frac{1}{2}
\sum_{\bm{k}}
\sum_{\ell=1}^{4}
\varepsilon_{\ell}(\bm{k})
. 
\label{eq:deltaE_MF}
\end{align}
%%%%%%%%%%%%%%%
We numerically evaluate $\delta E_{\mathrm{MF}}$ 
as a function of $0\leq\zeta\leq\pi/2$ as shown in Fig.~\ref{fig:lt}(d) at $\theta=7\pi/9$.
We find that the stripe order ($\zeta=0$) is lower in energy than the vortex order ($\zeta=\pi/4$). 
However, the energy difference between the two is vanishingly small as $\theta$ 
increases to $\theta \gtrsim 5\pi/6$ as shown in Fig \ref{fig:lt}(e). 
Therefore, near $\theta\sim\pi$, 
the quantum corrections beyond the present treatment, 
such as the higher-order ring-exchange interaction we neglected in deriving the effective spin model, 
or the magnon-magnon interactions from the spin-wave theory 
can easily modify the types of magnetic order in the ground state.
Indeed, the system favors the vortex order in the large-$\theta$ region of the SSDMF solution. 

%%%%%%%%%%%%%%%%%%%%%%%%%%%%%%%%%%%%%%%
%%%%%%%%%%%%%%% section %%%%%%%%%%%%%%%
%%%%%%%%%%%%%%%%%%%%%%%%%%%%%%%%%%%%%%%
\section{Sine-square deformed mean field theory}
\label{sec:ssdmf}
In this section, we explain the details of the SSDMF calculation which we performed 
in deriving the phase diagram shown in Fig.~\ref{fig:phase}(a). 
To quantitatively evaluate the magnetic phase diagram, the SSDMF is so far the best unbiased method. 
Indeed, the cluster DMFT and the mean field solutions with periodic boundaries 
cannot capture the subtle differences between the energies of different magnetic structures of the long 
spatial period; no matter how carefully one chooses the size and the shape of the clusters, 
the solutions are biased in practice, and the energies of the candidate solutions  
suffer the inevitable mismatch of the period of the lattice and orders. 
As we explained and demonstrated in Ref.[\onlinecite{kawano2022prr}] the solutions obtained by SSDMF does not suffer such effect, even when the ordering period is several times larger than the cluster we use. 
\par
\begin{figure}[t]
\includegraphics[width=85mm]{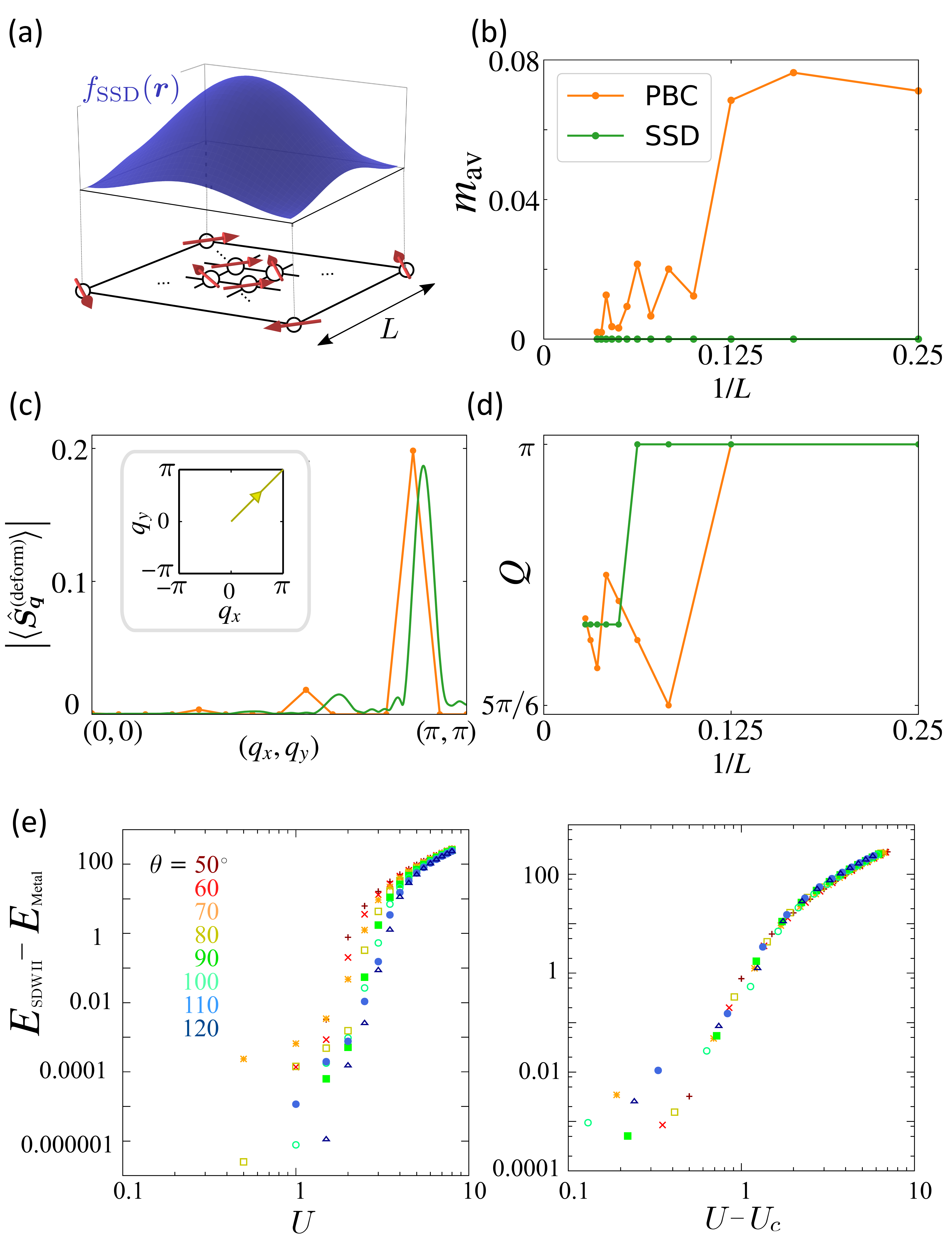}
\caption{
(a) Schematic illustration of the model with the SSD. 
(b) System-size dependence of the average magnetization for the AFM and metallic phase 
at $\theta=\pi/3$ and $U=0.5$.
(c) Structure factor as the function of $\bm{q}$ obtained from Eq.~(\ref{eq:deformed-fourier}) together with the PBC result
along the symmetric line in the Brillouin zone.
(d) System-size dependence of the peak position at $\theta=\pi/6$ and $U=3.0$.
(e) Energy difference between the SDWII and metallic phase of the same $\theta$ as a function of 
$U$ (left) and $U-U_c$ (right panel) for several choices of $\theta$, where $U_c$ is the phase boundary. 
}
\label{fig:ssdmf}
\end{figure}

We now briefly outline the SSDMF. 
The schematic illustration of the system with the SSD is shown in Fig.~\ref{fig:ssdmf}(a). 
We spatially modify the Hamiltonian by the sine-squared envelope function, 
%%%%%%%%%%%%%%%
\begin{align}
f_{\mathrm{SSD}}(\bm{r})
=
\frac{1}{2}
\left\{
1
+
\cos
\left(
\frac{\pi|\bm{r}|}{R}
\right)
\right\}
,
\end{align}
%%%%%%%%%%%%%%%
which has a straw-hat-like form, taking a maximum at the center of the 2D cluster 
(origin of the positional vector $\bm{r}$) 
with a radius $R=R_{0}+1/2$, where $R_{0}$ is the distance of the farthest site from the center~\cite{hotta2013prb}. 
Then, we perform a mean-field approximation as 
%%%%%%%%%%%%%%%
\begin{align}
&
\hat{\mathcal{H}}_{\mathrm{MF}}
=
E_{c}
\nonumber \\
&
-
t_{\mathrm{eff}}
\sum_{\bm{r}}
\sum_{\mu}
f_{\mathrm{SSD}}\left(
\bm{r}+\frac{\bm{e}_{\mu}}{2}
\right)
\left(
\hat{\bm{c}}_{\bm{r}+\bm{e}_{\mu}}^{\dagger}
\mathrm{e}^{i(\theta/2)\bm{n}_{\mu}\cdot\bm{\sigma}}
\hat{\bm{c}}_{\bm{r}}
+
\mathrm{h.c.}
\right)
\nonumber \\
&
+
U
\sum_{\bm{r}}
f_{\mathrm{SSD}}(\bm{r})
\left(
\frac{1}{2}
\left(
\braket{\hat{n}_{\bm{r}}}_{\mathrm{MF}}
-
1
\right)
\hat{n}_{\bm{r}}
-
2
\braket{\hat{\bm{S}}_{\bm{r}}}_{\mathrm{MF}}
\cdot
\hat{\bm{S}}_{\bm{r}}
\right)
.
\end{align}
%%%%%%%%%%%%%%%
where the first term is the constant given as 
%%%%%%%%%%%%%%%
\begin{align}
E_{c}
=
-
U
\sum_{\bm{r}}
f_{\mathrm{SSD}}(\bm{r})
\left(
\frac{1}{4}
\left(
\braket{\hat{n}_{\bm{r}}}_{\mathrm{MF}}^{2}
-
1
\right)
-
\braket{\hat{\bm{S}}_{\bm{r}}}_{\mathrm{MF}}^{2}
\right)
.
\end{align}
%%%%%%%%%%%%%%%
The mean fields $\braket{\hat{n}_{\bm{r}}}_{\mathrm{MF}}$ and $\braket{\hat{\bm{S}}_{\bm{r}}}_{\mathrm{MF}}$
are site-dependent and are determined self-consistently. 
\par
In finding the global minimum of energy, 
we examine several different types of initial values of the mean field 
by referring to the results of the RPA and Luttinger-Tisza method, 
and to the standard AFM order observed in the previous studies. 
To be more precise, 
the initial value of the particle density is set to $\braket{\hat{n}_{\bm{r}}}_{\mathrm{MF}}=1$, 
and we considered not only the single-$\bm{Q}$ state but the double-$\bm{Q}$ and higher-order-$\bm{Q}$ states
constructed by the combination of orders found in other approximations.
From the mean-field solution,
the ordering wave vector is extracted using the deformed Fourier transformation~\cite{kawano2022prr}
%%%%%%%%%%%%%%%
\begin{align}
\braket{\hat{\bm{S}}_{\bm{q}}^{(\mathrm{deform})}}_{\mathrm{MF}}
=
\frac
{\sum_{\bm{r}}f_{\mathrm{SSD}}(\bm{r})\braket{\hat{\bm{S}}_{\bm{r}}}_{\mathrm{MF}}\mathrm{e}^{-i\bm{q}\cdot\bm{r}}}
{\sum_{\bm{r}}f_{\mathrm{SSD}}(\bm{r})}
,
\label{eq:deformed-fourier}
\end{align}
%%%%%%%%%%%%%%%
where we take $\bm{q}=(q_{x},q_{y})$ as a continuous variable.
%%%%%%%%%%%%%%%%%%%%%%%%%%%%%%%%%%%%%%%%%%%%%%%%%%%%%%%%%%%%%%%%%%%%%%%%%%%%%%%%%%%%%%%%%%%%%%%%%%%%%%%%%%%%%%%%%%%%%%%%
%%%%%%%%%%%%%%%%%%%%%%%%%%%%%%%%%%%%%%%%%%%%%%%%%%%%%%%%%%%%%%%%%%%%%%%%%%%%%%%%%%%%%%%%%%%%%%%%%%%%%%%%%%%%%%%%%%%%%%%%
\par
We demonstrate that the SSD significantly suppresses the finite-size effect. 
Figure~\ref{fig:ssdmf}(b) shows the system size dependence of the averaged magnetization at $\theta=\pi/3$ and $U=0.5$.
For the PBC (standard mean-field solution without the SSD),
these quantities exhibit a substantial $L$-dependence,
and the magnetization predicts the AFM ordering for small $L$.
Contrastingly, in the SSDMF solutions, the average magnetizations have no detectable size dependence for $L\gtrsim 4$,
and they converge to the metallic state.
Figures~\ref{fig:ssdmf}(c) and \ref{fig:ssdmf}(d) show the spin structure factor and the size dependence of its peak position.
In Fig.~\ref{fig:ssdmf}(d) one finds that the ordering vector obtained for the PBC 
shows a large nonsystematic oscillations with its center off the true value to which the SSDMF one converges at $L\gtrsim 20$. 
This is because the true period of SDW has a mismatch with the cluster period 
and the moment is forced to form an artificial periodicity depending on $L$, 
which makes it difficult to give an accurate extrapolation to $L\rightarrow\infty$. 
The phase diagram in Fig.~\ref{fig:phase}(a) obtained by $L=28$ with the SSDMF 
is based on the data that are well converged, and are free of numerical artifacts 
from the size and shape of the cluster. 
\par
In deriving the phase diagram, the boundaries between different magnetic orders 
as a function of $\theta$ are the first-order transitions, which are obtained by the energy crossings 
of the two solutions. 
The metal-to-SDW transitions are more subtle; to evaluate the phase boundary, 
we measured the energy difference between the SDW phase ($E_{\mathrm{SDW}}$) 
and the metallic phase ($E_{\mathrm{metal}}$)  of the same $\theta$ at each $U$, where
we found that $E_{\mathrm{metal}}$ does not depend on $U$. 
As shown in Fig.~\ref{fig:ssdmf}(e), $E_{\mathrm{SDWII}}-E_{\mathrm{metal}}$ scales with $U-U_c$ 
for all different $\theta$ when we enter the SDWII phase. 
The values of $U_c$ plotted in the phase diagram are evaluated numerically as the ones that give the best collapse. 
The same analysis is applied to SDWII and III phases. 
%%%%%%%%%%%%%%%%%%%%%%%%%%%%%%%%%%%%%%%
%%%%%%%%%%%%%%% section %%%%%%%%%%%%%%%
%%%%%%%%%%%%%%%%%%%%%%%%%%%%%%%%%%%%%%%
\section{Density matrix embedding theory}
\label{sec:dmet}
The methods we applied so far all rely on the lowest order approximations about the correlation effect. 
The RPA and Luttinger-Tisza methods safely function in the weak and strong coupling regions, respectively, 
and they quantitatively agree well with the SSDMF phase diagram. 
Although we found previously that the SSDMF may accurately evaluate the Mott gap beyond the mean-field level~\cite{kawano2022prr}, 
how the effect of higher-order electronic correlations affects the other quantities is not fully clarified. 
\par
To support the SSDMF at moderately large $U$ where the mean-field approximation is the most fragile, 
we perform the DMET calculation. 
The DMET takes almost full account of the correlation effect, 
and reproduces the energy of the quantum Monte Carlo solutions with sufficient accuracy in the Hubbard models~\cite{knizia2012prl}. 
As we discuss shortly in Sec.\ref{sec:discussion}, the cluster-based methods are unsuccessful in evaluating 
the incommensurate phases. 
Although the DMET formulation makes use of the cluster, 
previous studies showed that the choice of cluster sizes and shapes does not influence the results~\cite{xavier2020prb}. 
Comparing the results of DMET and SSDMF will verify both of them in a complementary manner. 
\par
\begin{figure}[t]
\includegraphics[width=85mm]{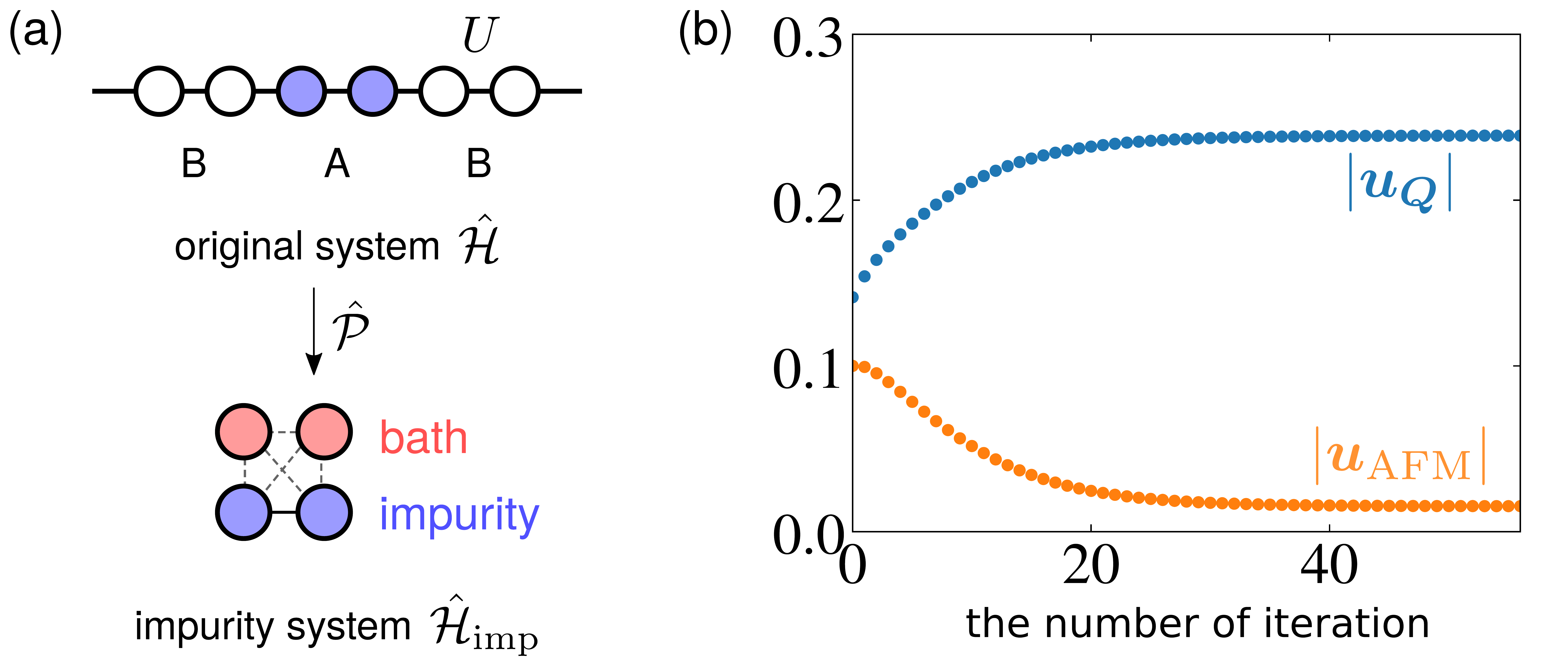}
\caption{Schematic illustration of the algorithm of the DMET.
(b) How the one-body potential develops during the DMET iteration for $\theta=\pi/6$ and $U=2.0$.}
\label{fig:dmet}
\end{figure}

In the following, we give a brief outline of the DMET shown schematically in Fig.~\ref{fig:dmet}(a), 
while the complete set of explanations and the benchmarks on 
several Hubbard models are given in Refs.\onlinecite{bulik2014prb,chen2014prb,zheng2016prb,zheng2017prb,kawano2020prb}.
We first divide the system into the small cluster A and the rest B, 
with $N_{\mathrm{A}}$ and $N_{\mathrm{B}}=N-N_{\mathrm{A}}$ sites ($N_A\ll N_B$) 
and prepare a reference Hamiltonian which is a one-body Hamiltonian. 
For the present purpose, we choose 
\begin{align}
\hat{\mathcal{H}}_{\mathrm{ref}}
&=
-t_{\mathrm{eff}}
\sum_{\bm{r}}
\sum_{\mu}
\left(
\hat{\bm{c}}_{\bm{r}+\bm{e}_{\mu}}^{\dagger}
\mathrm{e}^{i(\theta/2)\bm{n}_{\mu}\cdot\bm{\sigma}}
\hat{\bm{c}}_{\bm{r}}
+
\mathrm{h.c.}
\right)
\nonumber \\
&\hspace{20pt}
+
\sum_{\bm{r}}
\hat{\bm{c}}_{\bm{r}}^{\dagger}
\left(
u_{\bm{r}}^{0}
\sigma^{0}
+
\bm{u}_{\bm{r}}
\cdot
\bm{\sigma}
\right)
\hat{\bm{c}}_{\bm{r}}
,
\end{align}
%%%%%%%%%%%%%%%
where $u_{\bm{r}}^{0}$ and $\bm{u}_{\bm{r}}=(u_{\bm{r}}^{x},u_{\bm{r}}^{y},u_{\bm{r}}^{z})$ are the one-body potentials.
The ground state wave function of $\hat{\mathcal{H}}_{\mathrm{ref}}$ is 
Schmidt decomposed into subsystems A and B as 
%%%%%%%%%%%%%%%
\begin{equation}
\ket{\Psi}
=
\sum_{n=1}^{\chi}
\lambda_{n}(\Psi)
\ket{\Psi_{n}^{[\mathrm{A}]}}
\otimes
\ket{\Psi_{n}^{[\mathrm{B}]}}
,
\label{eq:schmidt}
\end{equation}
where $\chi$ is the dimension of basis in A, and by using this $\ket{\Psi}$, 
the true Hamiltonian $\hat{\mathcal{H}}$ is projected onto A by 
$\hat P= \hat 1_A \otimes \sum_n \ket{\Psi_{n}^{[\mathrm{B}]}}\bra{\Psi_{n}^{[\mathrm{B}]}}$ as 
$\hat{\mathcal{H}}_{\mathrm{imp}}=\hat P \hat{\mathcal H} \hat P$. 
Because of small $N_A$, the quantum many-body wave function $|\Phi_{\mathrm{imp}}\rangle$ 
is obtained exactly as the ground state of $\hat{\mathcal{H}}_{\mathrm{imp}}$. 
If one could properly choose a set of one-body potential, 
the local density matrix $\rho^{\mathrm{[A]}}={\rm Tr}_B |\Phi_{\mathrm{imp}}\rangle\langle\Phi_{\mathrm{imp}}|$ 
almost perfectly reproduces the local density matrix of the true ground state of $\hat{\mathcal H}$. 
If this is attained, the exact quantum many-body wave function is obtained locally in A as $|\Phi_{\mathrm{imp}}\rangle$. 
Therefore, the problem is reduced to finding optimal potential sets 
which is done in the iterative self-consistent process. 
\par
In the standard DMET calculation, 
one assumes that the one-body potential is defined in a unit of impurity cluster and is periodically 
repeated over the entire system. 
Since this construction is not suitable for long-period orders, 
we assume $u_{\bm{r}}^{0}=0$ and apply the following form; 
%%%%%%%%%%%%%%%
\begin{align}
\bm{u}_{\bm{r}}
&=
\bm{u}_{\mathrm{AFM}}
\cos\bm{Q}_{\pi}\cdot\bm{r}
\nonumber \\
&\hspace{20pt}
+
\mathrm{Re}[\bm{u}_{\bm{Q}}]
\cos\bm{Q}\cdot\bm{r}
-
\mathrm{Im}[\bm{u}_{\bm{Q}}]
\sin\bm{Q}\cdot\bm{r}
,
\end{align}
%%%%%%%%%%%%%%%
where $\bm{u}_{\mathrm{AFM}}$ and $\bm{u}_{\bm{Q}}$ are the potentials
that favor magnetic orders with $\bm{Q}_{\pi}=(\pi,\pi)$ and $\bm{Q}$, respectively. 
In iteratively preparing the potentials during the self-consistent DMET calculation, 
these potentials gradually develop, 
and one can examine which of the potentials the system favors. 
\par
We consider $12\times12$ lattice sites with the PBC and $N_{\mathrm{A}}=2\times2$ impurities. 
Although $N_{\mathrm{A}}$ does not need to match the periods introduced in $\bm{u}_{\bm{r}}$, 
the size of the whole lattice, where the potential is defined, is better 
consistent with the both periods that may appear in $\bm{u}_{\bm{r}}$; 
for the SDW order, the value of $\bm Q$ is chosen whose wave numbers are the submultiple of $L$. 
For the SDWI state, we adopt the wavenumber $Q_{\rm SDWI}$ which is observed in the RPA 
and choose $L$ to have $Q_{\rm SDWI}$ a submultiple, which also does not exclude the competing AFM. 
\par
We focus on the small-$\theta$ region since there is a contradiction between 
the SSDMF predicting the single-$\bm{Q}$ SDW order 
and the cluster DMFT supporting the AFM. 
The DMET will test which of the orders to be favored in the absence/presence of size effect/correlation. 
More importantly, this parameter region is possibly realized in real materials
such as Pb/Si(111) monolayer~\cite{tresca2018prl} and delafossite oxides~\cite{sunko2017nature}.
Figure~\ref{fig:dmet}(b) shows the iteration step dependence of 
$\bm{u}_{\mathrm{Q}_{\pi}}$ and $\bm{u}_{\mathrm{Q}_{\mathrm{SDWI}}}$ for $\theta=\pi/6$ and $U=2.0$. 
The potential for SDWI grows and overwhelms the AFM. 
While there still remains a small but finite $\bm{u}_\mathrm{AFM}$, 
this should be due to the higher-order harmonics other than
$\bm{u}_{\mathrm{Q}_{\mathrm{SDWI}}}$, which possibly comes from the inaccuracy of the choice of 
$\bm{u}_{\mathrm{Q}_{\mathrm{SDWI}}}$, since we do not know the true $\mathrm{Q}_{\mathrm{SDWI}}$ {\it a priori}. 
However, apart from this small misfit, the results support the SSDMF results. 
%%%%%%%%%%%%%%%%%%%%%%%%%%%%%%%%%%%%%%%%%%%%%%%%%%%%%%%%%%%%%%%%%%%%%%%%%%%%%%%%%%%%%%%%%%%%%%%%%%%%%%%%%%%%%%%%%%%%%%%%
%%%%%%%%%%%%%%%%%%%%%%%%%%%%%%%%%%%%%%%%%%%%%%%%%%%%%%%%%%%%%%%%%%%%%%%%%%%%%%%%%%%%%%%%%%%%%%%%%%%%%%%%%%%%%%%%%%%%%%%%

%%%%%%%%%%%%%%%%%%%%%%%%%%%%%%%%%%%%%%%%%%%%%%%%%%%%%%%%%%%%%%%%%%%%%%%%%%%%%%%%%%%%%%%%%%%%%%%%%%%%%%%%%%%%%%%%%%%%%%%%
\begin{figure}[t]
\includegraphics[width=85mm]{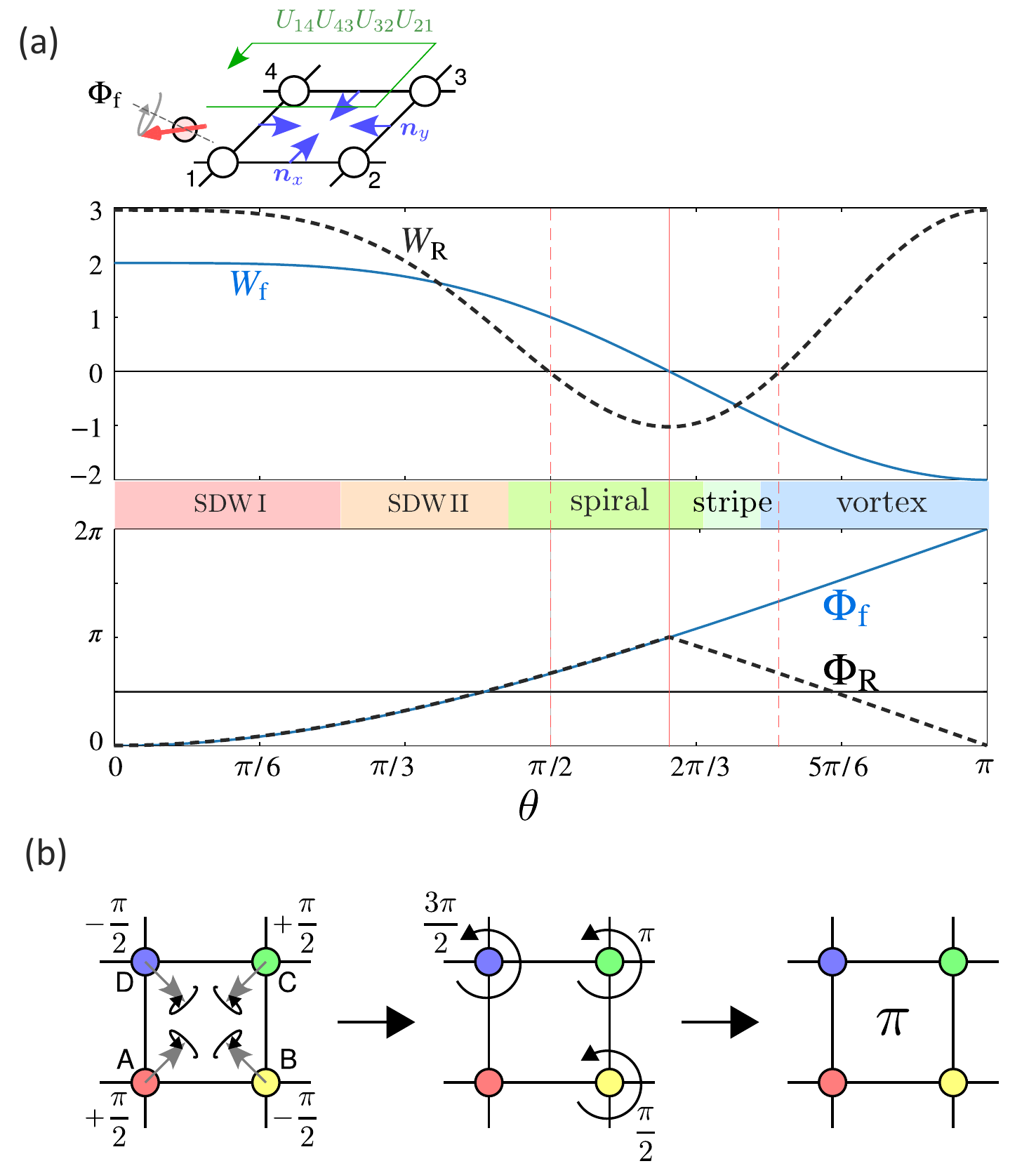}
\caption{
(a) Gauge-invariant quantities, $W_{\mathrm{f}}$, $\Phi_{\mathrm{f}}$, $W_{\mathrm{R}}$, $\Phi_{\mathrm{R}}$,
as functions of $\theta$, together with the magnetic phase diagram obtained by the SSDMF at $U\sim 8$. 
The red line indicates $\theta$ where $W_{\mathrm{f}}=0$ or $W_{\mathrm{R}}=0$.
The inset shows the four site cluster we consider.
(b) Schematic illustration of the local SU(2) gauge transformation
defined on four sublattices A (red), B (yellow), C (green), and D (blue).
The local SU(2) gauge transformation rotates the spin quantization axis by $\pm\pi/2$ about the gray arrows,
followed by the rotation about the $z$ axis.
}
\label{fig:wilson}
\end{figure}
%%%%%%%%%%%%%%%%%%%%%%%%%%%%%%%%%%%%%%%%%%%%%%%%%%%%%%%%%%%%%%%%%%%%%%%%%%%%%%%%%%%%%%%%%%%%%%%%%%%%%%%%%%%%%%%%%%%%%%%%
%%%%%%%%%%%%%%%%%%%%%%%%%%%%%%%%%%%%%%%
%%%%%%%%%%%%%%% section %%%%%%%%%%%%%%%
%%%%%%%%%%%%%%%%%%%%%%%%%%%%%%%%%%%%%%%
\section{Wilson loop}
\label{sec:wilson}
Part of the phase boundaries in Fig.~\ref{fig:phase}(a) can be explained analytically 
by using a gauge-invariant quantity called Wilson loop.
The concept of the Wilson loop was developed for the lattice gauge theory in high energy physics~\cite{rothe2012lattice}.
It is one of the fundamental gauge-invariant observables 
and serves as an ``order parameter'' that distinguishes between confined and deconfined phases of quarks. 
In the condensed matter field, 
the lattice gauge theory is applied to the topological phases~\cite{fradkin2013field}, 
where the Wilson loop characterizes the fractionalization of spin degrees of freedom. 
Here, we propose another useful property of the Wilson loop.
\par
We consider a four-site cluster in the square lattice shown in Fig.~\ref{fig:wilson}(a).
For later convenience, we introduce the following notation,
%%%%%%%%%%%%%%%
\begin{equation}
U_{21}
=
U_{12}^{\dagger}
=
U_{43}^{\dagger}
=
U_{34}
=
\mathrm{e}^{i(\theta/2)\sigma^{y}}
,
\end{equation}
%%%%%%%%%%%%%%%
%%%%%%%%%%%%%%%
\begin{equation}
U_{32}
=
U_{23}^{\dagger}
=
U_{14}^{\dagger}
=
U_{41}
=
\mathrm{e}^{-i(\theta/2)\sigma^{x}}
,
\end{equation}
%%%%%%%%%%%%%%%
where $U_{ij}$ is the SU(2) gauge field from site $j$ to $i$ ($i,j=1,2,3,4$).
In hopping around this closed loop, 
the electron acquires a gauge $U_{14}U_{43}U_{32}U_{21}$
and the electron spin points in a direction different from the initial one. 
Since $U_{ij}$ is the element of SU(2),
so as their product, which is rewritten as 
%%%%%%%%%%%%%%%
\begin{equation}
\mathrm{e}^{i(\Phi_{\mathrm{f}}/2)\bm{m}_{1}\cdot\bm{\sigma}}
=
U_{14}U_{43}U_{32}U_{21}
,
\label{eq:wilson_op}
\end{equation}
%%%%%%%%%%%%%%%
where $\Phi_{\mathrm{f}}\in[0,2\pi]$ denotes the rotation angle 
and the three-dimensional unit vector $\bm{m}_{1}\in\mathbb{R}^{3}$ determines the direction of the rotation axis. 
Wilson loop $W_{\mathrm{f}}$ is defined by the trace of Eq. (\ref{eq:wilson_op}), 
%%%%%%%%%%%%%%%
\begin{equation}
W_{\mathrm{f}}
=
\mathrm{Tr}
\left[
U_{14}U_{43}U_{32}U_{21}
\right]
=
2
\cos
\left(
\frac{\Phi_{\mathrm{f}}}{2}
\right)
.
\label{eq:Wf}
\end{equation}
%%%%%%%%%%%%%%%
By calculating $U_{14}U_{43}U_{32}U_{21}$, we obtain~\cite{sun2017njp}
%%%%%%%%%%%%%%%
\begin{equation}
W_{\mathrm{f}}
=
2
\left(
1
-
2
\sin^{4}\frac{\theta}{2}
\right)
.
\label{eq:Wf=2(1-2sin4)}
\end{equation}
%%%%%%%%%%%%%%%
Let us consider the local SU(2) gauge transformation, 
which rotates the spin quantization axis at site $i$ by $\phi_{i}$ about the unit vector $\bm{\eta}_{i}$ ($i=1,2,3,4$). 
The SU(2) gauge field $U_{i,j}$ is transformed as 
%%%%%%%%%%%%%%%
\begin{equation}
U_{ij}
\to
\mathrm{e}^{i(\phi_{i}/2)\bm{\eta}_{i}\cdot\bm{\sigma}}
U_{ij}
\mathrm{e}^{-i(\phi_{j}/2)\bm{\eta}_{j}\cdot\bm{\sigma}}
.
\label{eq:su2gauge_tr}
\end{equation}
%%%%%%%%%%%%%%%
From Eqs.~(\ref{eq:wilson_op}),~(\ref{eq:Wf}) and~(\ref{eq:su2gauge_tr}), 
one finds $W_{\mathrm{f}}\to W_{\mathrm{f}}$ and $\Phi_{\mathrm{f}}\to \Phi_{\mathrm{f}}$. 
Namely, the Wilson loop is gauge invariant. 
\par
One can apply the same argument for the effective spin Hamiltonian~(\ref{eq:Hspin-R}) 
in the strong-coupling limit~\cite{sun2017njp}, 
where the effect of SOC is described by the three-dimensional rotation matrix.
We consider the multiplication of the rotation matrices along the closed loop $1\to2\to3\to4$,
%%%%%%%%%%%%%%%
\begin{equation}
R(\bm{m}_{\mathrm{R}},\Phi_{\mathrm{R}})
=
R^{x}(\theta)
R^{y}(-\theta)
R^{x}(-\theta)
R^{y}(\theta)
. 
\end{equation}
%%%%%%%%%%%%%%%
It rotates the spin by $\Phi_{\mathrm{R}}$ about the unit vector $\bm{m}_{\mathrm{R}}$.
We introduce the $R$-matrix Wilson loop as 
%%%%%%%%%%%%%%%
\begin{equation}
W_{\mathrm{R}}
=
\mathrm{tr}
\left[
R(\bm{m}_{\mathrm{R}},\Phi_{\mathrm{R}})
\right]
=
1
+
2\cos\Phi_{\mathrm{R}}
, 
\end{equation}
%%%%%%%%%%%%%%%
and after the straightforward calculation, we have~\cite{sun2017njp}
%%%%%%%%%%%%%%%
\begin{equation}
W_{\mathrm{R}}
=
4
\left(
1
-
2
\sin^{4}\frac{\theta}{2}
\right)^{2}
-
1
.
\end{equation}
%%%%%%%%%%%%%%%
The rotation angle $\Phi_{\mathrm{R}}$ and $R$-matrix Wilson loop $W_{\mathrm{R}}$ are 
both gauge invariant. 
\par
Figure~\ref{fig:wilson}(a) shows the gauge-invariant quantities $W_{\mathrm{f}}$, $\Phi_{\mathrm{f}}$, $W_{\mathrm{R}}$, $\Phi_{\mathrm{R}}$,
as functions of $\theta$, together with the $U\sim 8$ phases extracted from the phase diagram. 
We find that except for the phase boundary that separates SDWI and SDWII phases,
the phase boundaries are close to $\theta$ where $W_{\mathrm{f}}=0$ or $W_{\mathrm{R}}=0$.
This result implies that there is a relationship between the phase boundaries and the gauge-invariant quantities. 
The SU(2) gauge transformation transforms the representation of the SOC Hamiltonian, 
whereas the magnetic phase boundaries may remain unchanged by this transformation 
since this transformation simply rotates the spin quantization axis. 
Therefore, the phase boundaries and the Wilson-loop, both being gauge invariant shall have some relationships. 
\par
Finally, we introduce a gauge transformation that transforms the SOC to the $\pi$-flux. 
As we saw in Fig.~\ref{fig:wilson}(a), 
the Wilson loops become $W_{\mathrm{f}}<0$ for large $\theta$, which may indicate the presence of a $\pi$-flux inside this plaquette
since the insertion of the $\pi$-flux changes the sign of $U_{14}U_{43}U_{32}U_{21}$ and thus changes the sign of $W_{\mathrm{f}}$.
In addition, at $\theta=\pi$,
we have $W_{f}=-2$ and $\Phi_{\mathrm{f}}=2\pi$ and become path-independent.
Therefore, the system at $\theta =\pi$ should be equal to the SU(2) symmetric Hubbard model with the $\pi$-flux,
which is similar to the spin-orbital quantum liquid in $\alpha$-ZrCl$_{3}$ and other spin-orbital models
where the system can be mapped to the SU(4) symmetric model with $\pi$-flux at strong SOC~\cite{yamada2018prl,yamada2021prb}.
Based on these considerations, we construct the unitary operator
that transforms the strong Rashba SOC with $W_{\mathrm{f}}<0$ to the weak antisymmetric SOC with $W_{\mathrm{f}}>0$ and with the $\pi$-flux. 
This operator, if exists, does not change the Hamiltonian at $W_{\mathrm{f}}=0$. 
\par
We first divide the system into four sublattices $X=$A,B,C,D as shown in Fig. \ref{fig:wilson}(b),
and consider the following local SU(2) gauge transformation 
%%%%%%%%%%%%%%%
\begin{equation}
\hat{\mathcal{U}}_{\mathrm{SU(2)}}
=
\hat{\mathcal{U}}_{\mathrm{SU(2)}}^{(\mathrm{A})}
\otimes
\hat{\mathcal{U}}_{\mathrm{SU(2)}}^{(\mathrm{B})}
\otimes
\hat{\mathcal{U}}_{\mathrm{SU(2)}}^{(\mathrm{C})}
\otimes
\hat{\mathcal{U}}_{\mathrm{SU(2)}}^{(\mathrm{D})}
,
\label{eq:U_su2_pi}
\end{equation}
where $\hat{\mathcal{U}}_{\mathrm{SU(2)}}^{(\mathrm{X})}$ acts only 
on sublattice $X$ and is defined as 
\begin{align}
\hat{\mathcal{U}}_{\mathrm{SU(2)}}^{(\mathrm{A})}
&=
\bigotimes_{\bm{r}\in\mathcal{I}_{\mathrm{A}}}
\exp
\left(
i
\frac{\pi}{2}
\frac{\bm{e}^{x}+\bm{e}^{y}}{\sqrt{2}}
\cdot
\hat{\bm{S}}_{\bm{r}}
\right)
,
\\
\hat{\mathcal{U}}_{\mathrm{SU(2)}}^{(\mathrm{B})}
&=
\bigotimes_{\bm{r}\in\mathcal{I}_{\mathrm{B}}}
\exp
\left(
i
\frac{\pi}{2}
\hat{S}_{\bm{r}}^{z}
\right)
\exp
\left(
-i
\frac{\pi}{2}
\frac{-\bm{e}^{x}+\bm{e}^{y}}{\sqrt{2}}
\cdot
\hat{\bm{S}}_{\bm{r}}
\right)
,
\\
\hat{\mathcal{U}}_{\mathrm{SU(2)}}^{(\mathrm{C})}
&=
\bigotimes_{\bm{r}\in\mathcal{I}_{\mathrm{C}}}
\exp
\left(
i
\pi
\hat{S}_{\bm{r}}^{z}
\right)
\exp
\left(
i
\frac{\pi}{2}
\frac{-\bm{e}^{x}-\bm{e}^{y}}{\sqrt{2}}
\cdot
\hat{\bm{S}}_{\bm{r}}
\right)
,
\\
\hat{\mathcal{U}}_{\mathrm{SU(2)}}^{(\mathrm{D})}
&=
\bigotimes_{\bm{r}\in\mathcal{I}_{\mathrm{D}}}
\exp
\left(
i
\frac{3\pi}{2}
\hat{S}_{\bm{r}}^{z}
\right)
\exp
\left(
-i
\frac{\pi}{2}
\frac{\bm{e}^{x}-\bm{e}^{y}}{\sqrt{2}}
\cdot
\hat{\bm{S}}_{\bm{r}}
\right)
.
\end{align}
%%%%%%%%%%%%%%%
Here $I_{X}$ denotes the set of sites on sublattice $X$. 
We apply this unitary operation to our Hamiltonian.
The on-site interaction term is obviously invariant. 
Because of the translational invariance, 
we only need to consider the transformation of the eight SU(2) gauge fields
$U_{\mathrm{XY}}=U_{i,j}$ ($i\in\mathcal{I}_{\mathrm{X}}$ and $j\in\mathcal{I}_{\mathrm{Y}}$).
For example,
$U_{\mathrm{BA}}$ is transformed as
%%%%%%%%%%%%%
\begin{align}
U_{\mathrm{BA}}
&\to
\mathrm{e}^{i \frac{\pi}{4} \sigma^{z}}
\mathrm{e}^{i \frac{\pi}{4}
   \frac{\bm{e}^{x}-\bm{e}^{y}}{\sqrt{2}}
   \cdot
   \bm{\sigma}}
\mathrm{e}^{i \frac{\theta}{2} \sigma^{y} }
\mathrm{e}^{i \frac{\pi}{4}
   \frac{-\bm{e}^{x}-\bm{e}^{y}}{\sqrt{2}}
   \cdot
   \bm{\sigma} }
\nonumber \\
&=
\exp
\left(
i
\frac{\pi-\theta}{2}
\frac{-\bm{e}^{x}-\bm{e}^{y}+\sqrt{2}\bm{e}^{z}}{2}
\cdot
\bm{\sigma}
\right)
.
\end{align}
%%%%%%%%%%%%%%%
Therefore,
the local SU(2) gauge transformation (\ref{eq:U_su2_pi}) induces $\theta\to\pi-\theta$. 
The direction of the unit vector $\bm{n}_{\mu}$ ($\mu=x,y$) also changes by this transformation.
Other SU(2) gauge fields are transformed as
%%%%%%%%%%%%%%%
\begin{align}
U_{\mathrm{CB}}
&\to
\exp
\left(
i
\frac{\pi-\theta}{2}
\frac{-\bm{e}^{x}-\bm{e}^{y}-\sqrt{2}\bm{e}^{z}}{2}
\cdot
\bm{\sigma}
\right)
,
\nonumber \\
U_{\mathrm{CD}}
&\to
\exp
\left(
i
\frac{\pi-\theta}{2}
\frac{\bm{e}^{x}+\bm{e}^{y}-\sqrt{2}\bm{e}^{z}}{2}
\cdot
\bm{\sigma}
\right)
, 
\end{align}
%%%%%%%%%%%%%%%
and so on. 
The transformation $\theta\to\pi-\theta$ means that the hopping amplitude $t=t_{\mathrm{eff}}\cos(\theta/2)$ 
and the strength of Rashba SOC $\lambda=t_{\mathrm{eff}}\sin(\theta/2)$ are exchanged.
Notice that some of the SU(2) gauge fields change their sign,
indicating the insertion of the $\pi$-flux in the system.
In particular, the large Rashba SOC ($\theta\sim\pi$) is transformed to weak SOC ($\theta\sim0$) Hubbard model
with $\pi$-flux.
We will use this finding in the next section in discussing the absence of spin liquid phase.
%%%%%%%%%%%%%%%%%%%%%%%%%%%%%%%%%%%%%%%
%%%%%%%%%%%%%%% section %%%%%%%%%%%%%%%
%%%%%%%%%%%%%%%%%%%%%%%%%%%%%%%%%%%%%%%
%%%%%%%%%%%%%%%%%%%%%%%%%%%%%%%%%%%%%%%%%%%%%%%%%%%%%%%%%%%%%%%%%%%%%%%%%%%%%%%%%%%%%%%%%%%%%%%%%%%%%%%%%%%%%%%%%%%%%%%%
\begin{figure}[t]
\includegraphics[width=85mm]{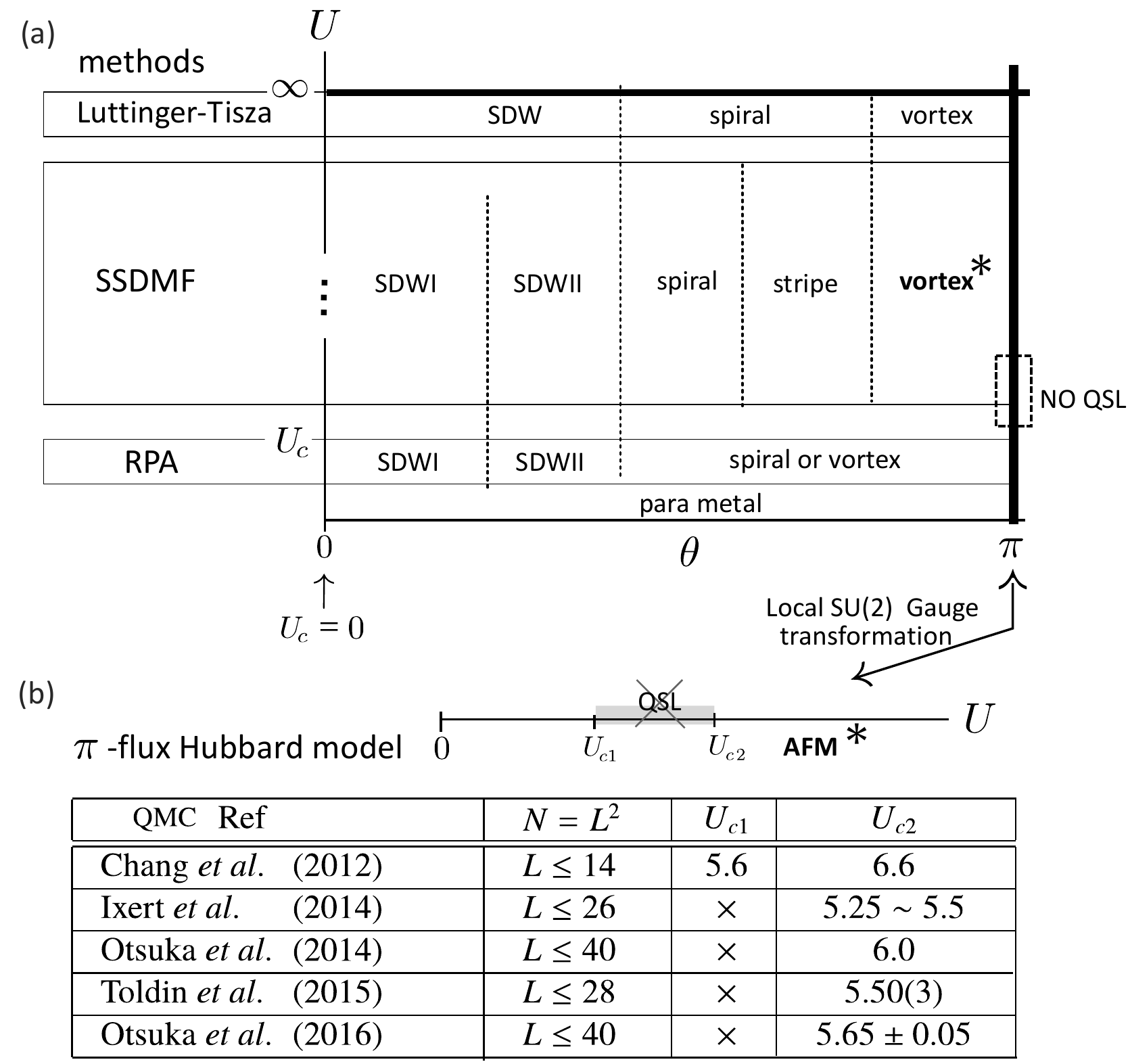}
\caption{
(a) Relationships between the approximations used in this study. 
The Luttinger-Tisza method is verified at around $U=\infty$, and the RPA at $U\lesssim U_c$, 
which give consistent results with the SSDMF results. 
(b) Summary of the results of the previous calculations on the $\pi$-flux Hubbard model 
which is identical to the $\theta=\pi$ case of our model after we perform the local SU(2) gauge transformation. 
The AFM(market with $*$) of the $\pi$-flux Hubbard model is the vortex phase of our model. 
The large-scale QMC results of larger sizes all predict the absence of the QSL phase ($U_{c1}$ being absent)
\cite{ixert2014prb,otsuka2014jpscp,toldin2015prb,otsuka2016prx}, 
showing that Chang {\it et al.}\cite{chang2012prl} using $L\le 14$ suffer a numerical finite size effect.
}
\label{fig:reschart}
\end{figure}
%%%%%%%%%%%%%%%%%%%%%%%%%%%%%%%%%%%%%%%%%%%%%%%%%%%%%%%%%%%%%%%%%%%%%%%%%%%%%%%%%%%%%%%%%%%%%%%%%%%%%%%%%%%%%%%%%%%%%%%%
\section{Discussion}
\label{sec:discussion}
We show in Fig.~\ref{fig:reschart}(a) the schematic chart showing which part of the parameter regions 
the methods we chose can be safely applied, and their conclusions about the ground state. 
\par 
First, we develop some discussions and remarks on the related theoretical studies. 
The phase diagram, or equivalently Fig.~\ref{fig:reschart}(a), contradicts the previously reported ones. 
Let us explain the overall phase diagram from the cluster DMFT study by Zhang, {\it et. al.}~\cite{zhang2015njp}; 
first of all, at $\theta=0$, the metal-insulator transition takes place at $U/t\sim 2$. 
There is a robust AFM order with $\bm{Q}=(\pi,\pi)$ 
up to $\theta \le \pi/3$, which transforms directly to the spiral phase, 
and at around $\theta\sim \pi/2$, the stripe phase appears, and at 
$\theta \gtrsim 2\pi/3$, they find the vortex phase. 
The nonmagnetic insulating phase is added at the metal-to-insulator transition 
point $U/t\sim 5$ and $\theta\sim \pi$. 
\par
Unfortunately, the following two issues can be trivially concluded: 
firstly, their onset value to the AFM phase $U_c/t \sim 2$ at $\theta=0$ is improper. 
Since the system suffers a perfect nesting due to the square shape of the Fermi surface, 
the insulating AFM phase starts immediately at $U\ne 0$ at $T=0$. 
Although having a nonzero $U_c$ at finite temperature is natural, 
the temperature $k_BT=0.05$ they adopted is still very low compared to the bandwidth 
and does not require such large $U_c$. 
Secondly, the nonmagnetic insulating phase at $\theta=\pi$ can be proved to be absent, 
which we explain shortly in this section.
\par 
A more serious difference is the lack of SDW phases in their phase diagram. 
We consider that the DMFT using the $4\times4$ cluster could not capture 
the long spatial period of incommensurate SDW orderings, which often happens in cluster-based methods. 
Although the spiral, stripe, and vortex phases may roughly agree with our results, 
they appear at $U/t\gtrsim 4-6$, apparently underestimating their stability compared to 
our RPA analysis yielding $U/t\gtrsim 2-3$.
The same results also hold for the four-site cluster model~\cite{brosco2018prb}.
Again, although $U_c$ tends to take larger values at finite temperature, 
they report the value three times larger than the RPA result ($U_c \sim2.0$) which is given at the same $k_BT=0.05$
and is rather close to that of the nonmagnetic metal-insulator transition~\cite{brosco2020prb}.
\par
We now prove the absence of spin liquid phase which Zhang {\it et.al.} claims in their phase diagram.
The three potential platforms of QSL known so far are the 
quantum spins in geometrically frustrated lattices~\cite{balents2010nature}, 
systems with frustrated exchanges in a nonfrustrated lattice such as Kitaev materials~\cite{kitaev2006,jackeli2009,chaloupka2010,singh2010,singh2012,plumb2014,kubota2015},
and the quantum many-body systems with large internal degrees of freedom represented by SU(N) spin systems with $N>2$~\cite{wang2009prb,corboz2011prl,corboz2012prx}. 
The essential features that are common to these three 
are some sort of competition or frustration and the enhanced quantum fluctuations. 
On the top of them, Meng and others proposed that the Dirac fermion systems can be another possible platform of QSL 
based on their QMC results on the honeycomb-lattice Hubbard model~\cite{meng2010nature}. 
They claimed that while the frustration is lacking, the small coordination number of the honeycomb lattice 
may enhance the quantum fluctuations. 
The later studies based on the QMC and other methods with careful finite-size extrapolation concluded that the QSL is absent~\cite{sorella2012sr,hassan2013prl,assaad2013prx,chen2014prb,toldin2015prb,otsuka2016prx}.
Subsequent studies supported the absence of QSL for other Dirac fermion systems, e.g. 
the square-lattice Hubbard model with a $\pi$-flux~\cite{ixert2014prb,otsuka2014jpscp,toldin2015prb,otsuka2016prx}. 
\par
Compared to these Dirac systems, our Dirac fermions may seem to have some room for the QSL phase since 
there is frustration/competition between 
DM interactions and Ising-type exchange interactions, both being induced by the Rashba SOC. 
However, we can still exclude the possibility of QSL; 
we have shown in the previous section that the present model at $\theta=\pi$ 
is equivalent to the SU(2) symmetric Hubbard model with $\pi$-flux, 
based on the local gauge transformation 
(see Fig.~\ref{fig:reschart}(b)).
 $\pi$-flux square lattice model is {\it shown not to have} a QSL phase 
by at least three independent QMC results \cite{ixert2014prb,toldin2015prb,otsuka2016prx}, 
except for oldest work~\cite{chang2012prl}, 
and instead, it shows a direct transition to the antiferromagnetically ordered state at $U_c\sim 6$. 
Since the local gauge transformation only varies the direction of spin-quantization axes 
and does not affect the nature of each phase, 
the AFM is transformed back to the other magnetic ordering. 
These considerations conclude that the nonmagnetic insulating phase in Ref.\onlinecite{zhang2015njp} is 
another artifact of the cluster-based calculation.
This situmation is  in contrast to $\alpha$-ZrCl$_{3}$ where
the emergent SU(4) symmetry leads to the spin-orbital liquid~\cite{yamada2018prl}.
\par
We finally remark on the difference between our SSDMF results and the previous standard mean-field result 
by Min\'{a}\v{r} and Gr\'{e}maud~\cite{minar2013prb}.
They performed the mean-field calculation with the periodic boundary condition (PBC) at a finite temperature 
and showed that for any $\theta$,
the system first enters the antiferromagnetically-ordered phase from the metallic one.
The recent paper by Kennedy and others also reported the AFM phase at small $\theta$~\cite{kennedy2022arxiv}. 
We have previously proved in a similar context~\cite{kawano2022prr} 
that these standard mean-field calculations give artificially stable solutions about CDW, 
incommensurate SDW~\cite{park2020prr} 
when the periodic boundary condition combined with the finite size of the unit cell 
restricts the types of mean-field solutions in advance. 
These are one of the numerical difficulties that we have mentioned in the introduction 
as a lack of appropriate theoretical tools to capture the large-scale structure 
in the present phase diagram, which was indeed not found for a long time. 
\par
Let us finally explain how the phase diagram in Fig.~\ref{fig:phase} (a) is safely concluded 
using Fig.~\ref{fig:reschart}.
Firstly, the QSL phases in the previous DMFT result is proved to be a numerical artifact,
once we accept the reliability of the four recent QMC works on the $\pi$-flux Hubbard model~\cite{ixert2014prb,otsuka2014jpscp,toldin2015prb,otsuka2016prx}. 
Secondly, the SSDMF is shown to safely capture the incommensurate phases if is present 
(see the size dependence in  Fig.~\ref{fig:ssdmf}), 
and indeed, the three SDW phases with incommensurate wave vectors appear in the phase diagram. 
As a third step, we checked both the quantitative and qualitative consistency of the SSDMF phase diagram 
with the RPA in the weak $U$ region (reliable in the weak-coupling phase) 
and with the Luttinger-Tisza method (reliable at large $U$ limit) in the large $U$ region. 
Finally, we confirmed that the SDWI phase which is replacing the AFM in the DMFT,
is energetically stable by using the DMET analysis. 
The paramagnetic-to-magnetic phase boundary can shift to higher $U$ when the higher-order correlation effect 
is taken account of. However, the representative phases we proposed, their origin, and their $\theta$-dependence 
is safely concluded as the basic nature of the model. 
\par
The four methods we applied can capture the incommensurate orderings if present. 
However, all unbiased quantum many-body numerical solvers using finite size clusters available so far cannot attack this problem. 
For example, the QMC has a sign problem, and the maximum size of density matrix renormalization group methods 
is much smaller than the size required, $L\times L$ with $L\gtrsim 100$ suggested in Fig.~\ref{fig:ssdmf}(d). 
Even in such cases, the present study demonstrates that the combination of methods can solve the issue. 
\vspace{5mm}
\\
%%%%%%%%%%%%%%%%%
\section{Summary}
\label{sec:summary} 
We studied the Mott Hubbard Hamiltonian having a Rashba-type of antisymmetric SOC 
and clarified the nature of the whole ground state. 
In particular, from weak to intermediate SOC regime, 
we find a transition from a metal to three different types of incommensurate SDW's. 
Such phases with large-scale spatial structures had been elusive for the models of 
strongly correlated electrons because of the lack of appropriate numerical solvers 
that can describe arbitrary types of magnetic orderings without bias. 
We applied the SSDMF which we developed recently, and combining it with other methods, 
demonsrtated that it works efficiently and reliably to clarify the ground states 
with many competing magnetic structures. 
\par
We have also clarified two different mechanisms of metal-to-magnetic phase transitions 
characteristic of antisymmetric SOC systems; 
since in these systems, the energy bands split by SOC because of the lack of inversion symmetry, 
the Fermi surface nesting instability works differently from the ordinary metals: 
it takes place between the Fermi surface that carries opposite spins, 
which we call ``spin pairwise nesting". Since such nesting occurs generally for incommensurate 
wave numbers, the long-period SDW appears. 
\par
Other interesting features of the present system appear for large SOC, the model hosts
spiral, stripe, and vortex phases of periods of two-lattice spacing, 
with magnetic moments of the large and same amplitudes, rotating in space. 
These phases appear just above the metal-insulator transition. 
The reason why they are easily stabilized by a relatively weak Coulomb interaction 
is ascribed to the four Dirac points with the same velocities in the time-reversal symmetric 
points (origin and edges of the Brillouin zone), located near the Fermi level. 
By nesting the whole Dirac cones by a wavelength $\pi$, the bandgap opens. 
Since the small density of states makes the Dirac systems generally stable against perturbation, 
this kind of phenomenon, driving the system to these magnets, 
is a remarkable feature of the antisymmetric Rashba SOC on a square lattice. 
\par
In the final part of the paper, we studied the role of a local gauge transformation. 
For example, some of the phase boundaries lie very close to the value of SOC 
at which the gauge-invariant Wilson-loop operator becomes zero. 
We also find the duality relationships between the strong SOC and weak-SOC parameter regions 
of the phase diagram, separated by that zero-Wilson-loop point, $\theta\sim 2\pi/3$;
the strong-SOC is equivalent to the weak-SOC phase with $\pi$-flux inserted in each plaquette.
In particular, the vortex order in the present model at $\theta=\pi$ (strong-SOC limit) 
can be mapped to the ordinary AFM state of the SU(2) symmetric $\pi$-flux Hubbard model 
by applying the local SU(2) gauge transformation. 
\par
We have confirmed the reliability of the SSDMF phase diagram by examining it using three other methods,
as well as proving that the counterpart QSL phases can hardly appear.
The phase diagram on the Rashba-SOC Hubbard model is almost fully updated.
%%%%%%%%%%%%%%%%%%%%%%%

%%%%%%%%%%%%%%%%%%%%%%%%%%%%%%%%%%%%%%%%%%%%%%%%
%%%%%%%%%%%%%%% acknowledgements %%%%%%%%%%%%%%%
%%%%%%%%%%%%%%%%%%%%%%%%%%%%%%%%%%%%%%%%%%%%%%%%
\section{Acknowledgements}
The authors thank Karlo Penc and Cristian D. Batista for fruitful discussion.
This work was supported by a Grant-in-Aid for Transformative Research Areas
``The Natural Laws of Extreme Universe— A New Paradigm for Spacetime and Matter from Quantum Information'' (No. 21H05191)
and JSPS KAKENHI (No.JP17K05533,21K03440).
M. K. was supported by JSPS Overseas Research Fellowship.

%%%%%%%%%%%%%%%%%%%%%%%%%%%%%%%%%%%%%%%%%%%%
%%%%%%%%%%%%%%% bibliography %%%%%%%%%%%%%%%
%%%%%%%%%%%%%%%%%%%%%%%%%%%%%%%%%%%%%%%%%%%%
\bibliography{biblio}
\bibliographystyle{apsrev4-1_mk}
\end{document}